%% file: ViBRWM_arxiv.tex
\documentclass{article}

\usepackage{vibrwm_preprint,times}

\usepackage{amsmath}
\usepackage{amssymb}
\usepackage{booktabs}
\usepackage{graphicx}
\usepackage{hyperref}
\usepackage{multirow}
\usepackage{placeins}
\usepackage{url}
\hypersetup{hidelinks,pdfstartpage={},pdfstartview={},
  pdftitle={ViBR-WM: Visual Bayesian Regression for World Modeling},
  pdfauthor={Jifan Li and Ning Ning}}

\newcommand{\methodname}{ViBR-WM}

\title{\raggedright ViBR-WM: Visual Bayesian Regression for World Modeling}

\author{Jifan Li \qquad Ning Ning\thanks{Corresponding author: \texttt{patning@tamu.edu}.}\\
Department of Statistics, Texas A\&M University\\
College Station, TX 77843}

\begin{document}
\raggedbottom
\maketitle
\pagestyle{plain}

\begin{abstract}
Modeling temporal dependence and uncertainty is central to forecasting
with world models.
The Visual Bayesian Regression World Model (\methodname{}) combines visual
features, physical histories and known covariates through interpretable regression,
within a modular architecture supporting trend, seasonal and cycle dynamics.
Visual compression reduces representation dimension, while Bayesian variable
selection reduces active regression dimension. Posterior prediction combines
forecasts across predictor subsets using their posterior probabilities as weights
and accounts for parameter uncertainty and future disturbances.
The model forecasts joint visual--physical states recursively and physical targets
directly. Across four forecasting tasks spanning object motion, vegetation
greenness and solar power, ViBR-WM achieves lower mean overall physical-target
error than Temporal Straightening, ConvLSTM, PredRNN and
SimVP on every task. Repeated fitting and resampling support these overall gains.
\end{abstract}

\section{Introduction}

World models learn how environments evolve in order to support prediction and planning \citep{ha2018worldmodels,hafner2025dreamerv3}. Rather than modeling this evolution directly in observation space, LeCun argues that useful world models should predict in an abstract representation space, motivating the Joint-Embedding Predictive Architecture (JEPA) \citep{lecun2022path}. The central intuition is that pixel-level prediction requires modeling large amounts of fine-grained visual detail that may be difficult to predict yet largely irrelevant for reasoning and action. JEPA instead encodes observations into latent representations and learns to predict the representations of future or unobserved states, allowing the model to focus its capacity on more predictable and semantically meaningful aspects of the environment.

DINO World Model (DINO-WM) provides a concrete realization of this principle for action-conditioned world modeling \citep{zhou2025dinowm}. Rather than learning a visual representation jointly with the dynamics model or reconstructing future images, DINO-WM uses a frozen DINOv2 image encoder to obtain spatial patch-level features and trains a predictor to forecast how these features evolve under candidate actions. This design is important because it shows that strong pretrained visual representations can serve directly as the state space of a world model: DINO-WM can learn dynamics from offline trajectories and perform zero-shot, task-agnostic planning toward image-specified goals without expert demonstrations, reward models, or inverse dynamics models. In this sense, DINO-WM closely follows the JEPA philosophy, while extending it into an actionable world model in which predicted latent dynamics can be used directly for planning.

However, frozen DINOv2 features are not optimized for planning: curved
latent trajectories can make Euclidean distance to a goal a poor measure of
progress along feasible transitions. Temporal Straightening (TS) addresses
this limitation within the JEPA framework by jointly learning a visual
representation and an action-conditioned predictor with a latent prediction
loss and an explicit trajectory-curvature penalty
\citep{wang2026straightening}.
In its DINO-based implementation, the backbone
remains frozen, while a trainable projector adapts its features jointly with
the predictor. Thus, TS extends DINO-WM's use of pretrained visual features
by learning a representation geometry suited to planning. TS improves gradient-based goal-reaching
over frozen-feature DINO-WM and prediction-only counterparts, making it a
strong JEPA-family comparator.

We propose the Visual Bayesian Regression World Model (ViBR-WM), which
combines prediction in visual representation space with Bayesian forecasting
of physical targets. For recursive forecasting, ViBR-WM builds directly on
TS: a TS-trained projector transforms frozen DINOv2 features, and principal
component analysis compresses them into visual coordinates. These fitted
representations remain fixed during Bayesian inference. ViBR-WM therefore
shares DINO-WM's pretrained visual backbone and JEPA's principle of
prediction in representation space, while its forecasting structure differs
from the jointly trained neural encoder--predictor of TS. It replaces the
neural transition predictor with Bayesian regression of changes in a joint
visual--physical state, conditioned on observed histories and known actions
or covariates. Horizon-specific regressions additionally support direct
physical-target prediction without intermediate rollout.

\begin{figure}[!t]
\centering
\includegraphics[width=\linewidth]{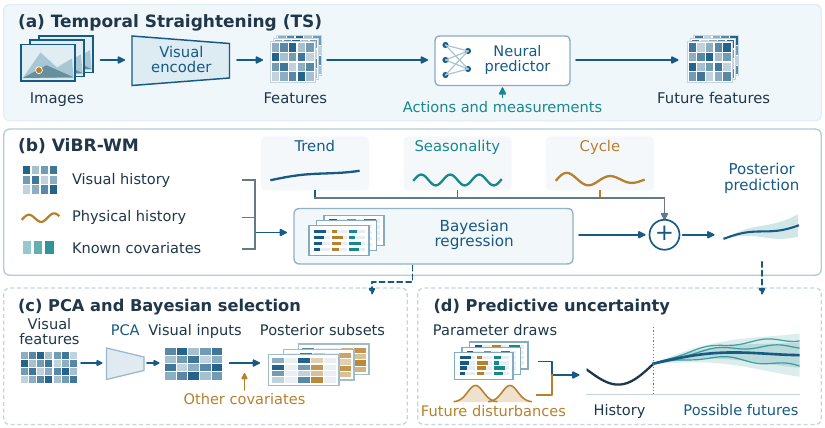}
\caption{Predictive architectures and Bayesian inference.
(a) TS predicts future features from encoded images, actions and observed
measurements. During training, prediction loss compares predicted features with
encoded future observations; the curvature penalty discourages changes in
direction between consecutive displacements along encoded trajectories of
observed images.
(b) ViBR-WM's trend, seasonal, cycle and regression contributions add before
posterior prediction. Dashed arrows expand the mechanisms below.
(c) Principal component analysis (PCA) compresses visual features, which join
other covariates as regression inputs. Bayesian selection assigns posterior
probabilities to predictor subsets. Posterior prediction combines the corresponding
predictive distributions using these probabilities as weights.
(d) Parameter uncertainty and future
disturbances yield distributions over possible futures.}
\label{fig:prediction-architecture}
\end{figure}

ViBR-WM's structural innovation integrates visual representations with an
additive Bayesian architecture comprising regression, trend, seasonal and
cycle modules (Figure~\ref{fig:prediction-architecture}). These modules
explicitly separate covariate effects, long-term change, seasonal recurrence
and stochastic oscillations; cycle damping allows disturbance effects to
decay. Many visual forecasting tasks record physical measurements or control
actions alongside images \citep{zhou2025dinowm,nie2023skippd}.
Many widely used forecasting methods make limited joint use of visual and numerical information.
ViBR-WM addresses this limitation by incorporating visual features, observed
physical histories and known covariates into its regression interface.
Bayesian variable selection reduces the active predictor set, while posterior
prediction averages over plausible subsets. This structure makes predictor
effects explicit and produces posterior predictive distributions that account
for parameter uncertainty and future disturbances.

We evaluate ViBR-WM on PointMaze and Wall, two benchmarks for visual world models \citep{zhou2025dinowm}, as well as on two real-world forecasting tasks: vegetation greenness forecasting using PhenoCam \citep{richardson2018phenocam} and solar-power forecasting using SKIPP'D \citep{nie2023skippd}. We compare ViBR-WM with TS and three widely adopted video-prediction methods: ConvLSTM \citep{shi2015convlstm}, PredRNN \citep{wang2017predrnn}, and SimVP \citep{gao2022simvp}. Relative to TS, ConvLSTM, PredRNN, and SimVP, respectively, ViBR-WM reduces mean overall normalized root mean squared error (NRMSE) by 16.19\%, 82.05\%, 50.89\%, and 60.98\% on PointMaze; 20.71\%, 62.79\%, 60.88\%, and 61.84\% on Wall; 74.20\%, 55.06\%, 57.52\%, and 50.37\% on PhenoCam; and 59.99\%, 57.02\%, 49.20\%, and 55.45\% on SKIPP'D. Repeated fitting and resampling analyses further support these overall error reductions.

Our contributions are threefold:
\begin{itemize}
    \item \textbf{Modular Bayesian architecture.}
    We introduce ViBR-WM, a visual world model that integrates learned
    visual representations with regression, trend, seasonal and cycle modules
    in an interpretable Bayesian architecture.
    \item \textbf{Forecasting with uncertainty quantification.}
    We develop recursive forecasting of joint visual--physical states
    and direct physical-target prediction, using Bayesian variable selection
    and posterior prediction to account for uncertainty in predictor subsets,
    coefficients and future disturbances.
    \item \textbf{Empirical validation across four tasks.}
    We demonstrate lower overall physical-target forecasting error than
    TS, ConvLSTM, PredRNN and SimVP across four tasks spanning object motion,
    vegetation dynamics and solar power, with repeated fitting and resampling
    analyses supporting these gains.
\end{itemize}

\section{Related work}

\paragraph{Visual world models and video prediction.}
Recurrent World Models learn compressed generative dynamics for control
\citep{ha2018worldmodels}; PlaNet is a reinforcement
learning agent that selects actions through latent-space planning
\citep{hafner2019planet}, while Dreamer agents learn action policies from imagined
trajectories \citep{hafner2021dreamerv2,hafner2025dreamerv3}.
IRIS models discrete visual tokens with an autoregressive Transformer,
and DIAMOND uses diffusion to predict future observations
\citep{micheli2023iris,alonso2024diamond}.
ConvLSTM's convolutional recurrence outperformed the then-leading ROVER
precipitation-nowcasting system \citep{shi2015convlstm}.
PredRNN's cross-layer spatiotemporal memory achieved state-of-the-art
prediction on three datasets \citep{wang2017predrnn}.
SimVP's convolutional encoder--translator--decoder achieved leading
performance on five benchmarks \citep{gao2022simvp}.
PredBench's 15-dataset, five-domain evaluation includes all three as widely
adopted methods \citep{wang2024predbench}.

\paragraph{Visual representations.}
Vision Transformers and self-supervised DINO features provide transferable
image representations \citep{dosovitskiy2021vit,caron2021dino,oquab2024dinov2}.
V-JEPA trains its video encoder and predictor to minimize the error in predicting
masked-region features from visible video content \citep{bardes2024revisiting}.
Earlier feature anticipation \citep{vondrick2016anticipating} and
DINO-based forecasting \citep{zhou2025dinowm,karypidis2025dinoforesight}
motivate forecasting directly in representation space.
Temporal Straightening encourages locally straighter latent trajectories during training
to improve latent-space planning \citep{wang2026straightening}.
JEPA-WM planning performance also depends on predictor architecture,
training objectives and planning algorithms \citep{terver2026jepawms}.
ViBR-WM uses fitted visual representations as inputs
to Bayesian forecasting models.

\paragraph{Regression and dynamical-system identification.}
Bayesian structural forecasting combines regression with dynamic components
\citep{scott2014predicting}; Bayesian variable selection
quantifies uncertainty about which predictors to include in the model
\citep{george1993variable}.
To model how physical states evolve over time, \citet{brunton2016sindy,brunton2016sindyc}
use sparse regression to select equation terms, while \citet{proctor2016dmdc}
use dynamic mode decomposition with control to fit linear state transitions.
\citet{lusch2018deep} learn coordinates that linearize dynamics.
\citet{champion2019coordinates} combine an autoencoder with sparse regression
to discover coordinates and equations.
\citet{gao2024bayesianautoencoder} give the autoencoder--regression framework
a Bayesian formulation.
Related Bayesian methods quantify uncertainty in term selection and coefficients
\citep{hirsh2022sparsifying}.

\paragraph{Predictive distributions and uncertainty.}
\citet{wan2026hauwm} encourage uncertainty to increase with
forecast horizon.
ViBR-WM represents stochastic futures through posterior
prediction over regression parameters and future disturbances. We evaluate
predictive distributions for physical targets with proper scoring rules
\citep{gneiting2007proper}, alongside point error and interval coverage.

\section{A Bayesian world model from visual and physical histories}

ViBR-WM combines visual representations with trend, seasonal, cycle
and regression modules (Figure~\ref{fig:prediction-architecture}).
Appendix~\ref{sec:world-model-inference} gives their joint observation and distinct
transition equations. The regression formulations below support recursive
and direct forecasting with parameter and disturbance uncertainty.

\subsection{Observations and state representation}

For recursive forecasting, $t$ indexes video frames. Let $o_t$ denote a
red--green--blue (RGB) image and $p_t\in\mathbb R^d$ the vector of $d$
physical measurements observed at that frame. DINOv2's frozen
small Vision Transformer with $14\times14$-pixel patches (ViT-S/14) encodes
$o_t$ as $e_t\in\mathbb{R}^{196\times384}$. A learned projector
$\mathcal{P}_{\phi}$, trained with the TS objective, produces a visual
patch field $z_t^v=\mathcal{P}_{\phi}(e_t)\in\mathbb{R}^{196\times8}$.
Each of the $14\times14=196$ locations therefore has eight feature channels.
The reduction map $\mathcal R_{\mathrm{train}}:\mathbb R^{196\times8}
\to\mathbb R^q$ converts this field to $q$ visual coordinates using two
successive applications of principal component analysis (PCA).
The first PCA compresses the channels at
each location. The resulting features are concatenated into one vector per
frame. A second PCA learns $q$ projection directions from changes between
frames $t$ and $t+\Delta$, where $\Delta$ is the number of frames per modeled
transition. Projecting each frame's vector onto these directions, then
centering and scaling the coordinates, gives
$\widetilde z_t^v=\mathcal R_{\mathrm{train}}(z_t^v)\in\mathbb R^q$.
Appendix~\ref{sec:world-model-inference} specifies its matrices and scales.

Let $\mu_p\in\mathbb R^d$ contain the means of the physical coordinates
over all training frames. Let $D_p\in\mathbb R^{d\times d}$ be the diagonal
matrix of their sample standard deviations. Their standardized
values and the joint state are
\[
 \bar p_t=D_p^{-1}(p_t-\mu_p),\qquad
 s_t=[(\widetilde z_t^v)^\top,\bar p_t^\top]^\top
 \in\mathbb R^m,\quad m=q+d.
\]

\subsection{Bayesian state-transition regression}

ViBR-WM gives users the flexibility to combine regression with trend, seasonal
and cycle modules according to the task. The purpose of including these modules
is to enhance interpretability by explicitly separating the contributions of
trend, seasonality and cycles from covariate effects in the forecast.
The trend module tracks an evolving baseline level and slope, the seasonal
module captures periodic recurrence, and the cycle module represents
oscillations that can decay after disturbances.
Regression gives covariates interpretable effects. In the full architecture,
the structural and regression modules jointly determine the response mean.
The trend, seasonal and cycle trajectories are inferred jointly with the
regression coefficients. Posterior forecasts combine the module contributions
and simulate disturbances to these trajectories, together with random
deviations of observations from
the resulting response mean (Appendix~\ref{sec:world-model-inference}).
Adding modules does not guarantee better forecasts: extra parameters can
overfit and increase prediction error.

The regression module models changes in the joint visual--physical state.
Let $a_t\in\mathbb R^{d_a}$ denote the action applied at frame $t$.
Its $d_a$ entries specify numerical control inputs, such as commands for
horizontal and vertical movement. The action block
$A_t=[a_t^\top,\ldots,a_{t+\Delta-1}^\top]^\top\in\mathbb{R}^{\Delta d_a}$
collects the actions governing the next transition. The prediction history is
\[
 \mathcal H_t=\bigl(s_t,s_{t-\Delta},\ldots,s_{t-(L-1)\Delta};\,
 A_t,A_{t-\Delta},\ldots,A_{t-(K-1)\Delta}\bigr).
\]
Here $L$ counts joint-state vectors from Section~3.1 at $\Delta$-frame intervals;
$K$ counts known action blocks spanning $\Delta$ consecutive frames each.
The two lengths can differ. The map $\chi$ gives regression inputs
$x_t\in\mathbb R^P$ with entries $x_{tk}=\chi_k(\mathcal H_t)$.
Each $\chi_k$ extracts a state/action value or calculates a
derived predictor. For example, multiplying a position coordinate by a mean
action coordinate within $A_t$ lets the action's effect depend on position.
Regressions using different input constructions are compared on validation data;
forecasting uses the lowest-error construction. Inclusion indicators below
select predictors within that construction.

Training window $i=1,\ldots,N$ has prediction origin $t_i$ and input
$x_i=x_{t_i}=\chi(\mathcal H_{t_i})$. Its target is the state change between
frames $t_i$ and $t_i+\Delta$ in the same video:
\begin{equation}
 r_i = s_{t_i+\Delta}-s_{t_i}\in\mathbb{R}^{m},\qquad
 r_i\mid x_i,\theta \sim
 \mathcal N_m\!\left(\mu_\theta(x_i),\Sigma\right).
\end{equation}
Here $\mu_\theta(x_i)\in\mathbb R^m$ is the conditional mean change and
$\Sigma\in\mathbb R^{m\times m}$ is its disturbance covariance.
Matrix $B\in\mathbb R^{m\times P}$ links $P$ input predictors to $m$
state-change coordinates. Its $j$th row $\beta_j^\top$ contains one coefficient
per predictor for predicting coordinate $j$. The conditional mean is
\[
 \mu_{\theta,j}(x)=\bar r_j+\beta_j^\top\{D_x^{-1}(x-\mu_x)-c_{x,j}\},
\]
where $\bar r_j\in\mathbb R$ is the training mean of the target changes $r_{ij}$.
Training predictor means $\mu_x$ and sample standard deviations on the diagonal
of $D_x$ define standardized inputs $D_x^{-1}(x_i-\mu_x)$.
The fit centers these inputs using their training-window mean vector $c_{x,j}\in\mathbb R^P$.
Coefficient $\beta_{jk}$ weights predictor $k$ for response $j$, conditional on
the other predictors.
The off-diagonal entries of $\Sigma$ capture dependence between prediction
errors in different visual and physical coordinates.

For predictor $k=1,\ldots,P$ and response $j$, the inclusion indicator
$\gamma_{jk}\sim\operatorname{Bernoulli}(\pi_{jk})$ sets $\beta_{jk}=0$
when $\gamma_{jk}=0$; $\pi_{jk}$ is its prior inclusion probability.
The regression coefficients $\beta_{jk}$ with $\gamma_{jk}=1$ jointly have a multivariate Gaussian prior with
mean vector zero, which shrinks their estimates toward zero. The disturbance
covariance matrix $\Sigma$ has an inverse-Wishart prior.
Combining these priors with the training
likelihood gives the joint posterior of $\theta=(B,\Sigma,\gamma)$, with
$\gamma\in\{0,1\}^{m\times P}$. Coefficients are constant over time.
The likelihood assumes conditionally
independent disturbances across windows; overlapping windows may have correlated errors.

\subsection{Recursive posterior prediction}

Each simulated trajectory $b$ starts at frame $t_0$ with observed history
$\mathcal H_{t_0}^{(b)}=\mathcal H_{t_0}$.
One joint posterior draw $\theta^{(b)}$ is used for all transitions along that trajectory
(Figure~\ref{fig:statistical-method}).
For $H$ transitions, define
$t_h=t_0+h\Delta$, $h=0,\ldots,H$. At step $h=0,\ldots,H-1$, form
$x_h^{(b)}=\chi(\mathcal H_{t_h}^{(b)})$, sample a state change $r_h^{(b)}$,
and add it to the current state:
\begin{equation}
 r_h^{(b)} \sim
 \mathcal{N}_m\!\left(\mu_{\theta^{(b)}}(x_h^{(b)}),\Sigma^{(b)}\right),\qquad
 s_{t_h+\Delta}^{(b)} = s_{t_h}^{(b)}+r_h^{(b)}.
\end{equation}
After each step, append the simulated state, retain the latest $L$ states,
and advance through the supplied action schedule by $\Delta$ frames.
Parameter draws
represent uncertainty in dynamics; stepwise Gaussian disturbances represent
future random variation.
Means and quantiles across paths give point forecasts and predictive intervals.
The standardized physical variables $\bar p_{t_h}^{(b)}$ in each sampled state
are converted back to their original measurement units using the training mean
$\mu_p$ and scales $D_p$ from Section~3.1:
$p_{t_h}^{(b)}=\mu_p+D_p\bar p_{t_h}^{(b)}$.

\input{figures/wall_forecasting_overview.tex}

\subsection{Direct physical-target forecasts}

For direct prediction, $t$ indexes observation times and $h\geq1$ counts
observation intervals ahead. The scalar target is $y_{t+h}\in\mathbb R$.
Let $\mathcal I_{t,h}$ contain image and numerical observations through $t$
and covariates known at $t$ for the future time $t+h$. The map $\chi_h$ concatenates
PCA-reduced DINOv2 features, numerical-history values and target-time covariates
into the regression input $x_{t,h}=\chi_h(\mathcal I_{t,h})\in\mathbb R^{P_h}$.
The target transform $u_{t+h}=(y_{t+h}-\mu_{y,h})/d_{y,h}$ uses the
training mean and standard deviation, or $(\mu_{y,h},d_{y,h})=(0,1)$
when retaining the original units.

For each $h$, training pairs $(x_{t_i,h},u_{t_i+h})$ link inputs at
$t_i$ to targets $h$ intervals later. With
$\theta_h=(\beta_h,\sigma_h^2,\gamma_h)$, the scalar regression is
\begin{equation}
 u_{t+h}\mid\mathcal I_{t,h},\theta_h\sim
 \mathcal N\!\left(\mu_{\theta_h}(x_{t,h}),\sigma_h^2\right),\qquad
 \mu_{\theta_h}(x)=\bar u_h+\beta_h^\top(x-c_{x,h}).
\end{equation}
Here $\beta_h\in\mathbb R^{P_h}$ contains the coefficients,
$\sigma_h^2$ is the disturbance variance, and $c_{x,h}\in\mathbb R^{P_h}$ and
$\bar u_h\in\mathbb R$ are
the training means of the predictor vector and transformed target.
The indicators $\gamma_h\in\{0,1\}^{P_h}$ follow the Bernoulli inclusion
rule in Section~3.2, setting $\beta_{hk}=0$ when $\gamma_{hk}=0$.
Each posterior draw retains
$P_{\mathrm{active},h}=\sum_{k=1}^{P_h}\gamma_{hk}$ predictors.
The predictive distribution averages the conditional Gaussian distributions over
posterior draws; selected predictor subsets can differ across targets
and horizons.

Posterior prediction draws $\theta_h$ from its joint posterior, then samples
$u_{t+h}$ from the Gaussian above and returns it to physical units as
$y_{t+h}=\mu_{y,h}+d_{y,h}u_{t+h}$. Each horizon is predicted directly,
without feeding back earlier predictions; the separate distributions do not
define a joint path. When computing these intervals, the fitted visual representation
and selected predictor construction are treated as fixed. Appendix~\ref{sec:world-model-inference}
gives posterior derivations and implementation details.

\section{Experimental design}

\subsection{Tasks and data splits}

Appendix~\ref{sec:representation} details the data-processing and evaluation protocol.
PointMaze and Wall use DINO-WM trajectories \citep{zhou2025dinowm}.
The targets are position and instantaneous velocity for PointMaze, and position for Wall.
Forecasts start at frame 10 and advance in five-frame steps. Predictions are evaluated
at frames 15, 25, 35 and 60 for PointMaze, and at frames 15, 25, 35 and 45 for Wall.
Each episode in the released data records a separate simulation run.
Within each task, episodes follow the same dynamics, with their own initial states and action sequences.
Episodes 0--15 form the training set for fitting models and preprocessing transformations,
16--19 form the validation set for selecting model configurations, and
100--149 form the test set of 50 trajectories.
For Wall, we correct position labels that do not match the objects shown in
the video by replaying the recorded actions in the original simulator.
The images and actions are unchanged. Appendices~\ref{sec:wall-native-final}
and~\ref{sec:pointmaze-native-final} give the Wall and PointMaze protocols and results.

PhenoCam uses version-3 imagery and vegetation-index data for the
Mead1 and Mead3 agricultural regions of interest (ROIs)
\citep{ballou2025phenocamimages,zimmerman2025phenocamgreenness,
richardson2018phenocam}, selecting one image per week, with pixels outside
the vegetation ROI replaced by a uniform gray background.
The target is midday green chromatic coordinate, Gcc = $G/(R+G+B)$,
where $R,G,B$ are red, green and blue intensities in the vegetation ROI.
Data from the years 2018--2020 are used for training, 2021 for validation and
2023 for testing. Forecasts are evaluated at 1, 4, 13 and 26 weeks ahead.

SKIPP'D provides synchronized one-minute fisheye images and photovoltaic
(PV) power \citep{nie2023skippd}: 2017 for training, 2018 for validation and
2019 for testing. Forecasts start at 3,124 different time points in the test data and
are evaluated 5, 15, 30 and 60 minutes later.
Appendix~\ref{sec:scalar-native-final} gives the PhenoCam and SKIPP'D specifications.

Task-specific mappings convert predicted images or feature embeddings into
position, velocity, Gcc or power for evaluation.
Appendices~\ref{sec:wall-native-final}--\ref{sec:scalar-native-final}
describe these mappings and the model implementations.

\subsection{Model fitting and selection}

For each task, reported errors average all ten repetitions of each method.
Each ConvLSTM, PredRNN and SimVP fit comprises 100 training epochs minimizing
RGB prediction loss. After training, test forecasts use the parameters from
the epoch with the lowest validation loss.
TS trains for 20 epochs using a prediction loss with a temporal-curvature penalty. Its PointMaze and Wall
forecasts use the parameters at the end of training; PhenoCam and SKIPP'D use
the parameters yielding the lowest total validation loss.
ViBR-WM estimates regression parameters by Markov chain Monte Carlo (MCMC)
sampling, for which neural-network training epochs are not applicable.
We therefore used roughly comparable wall-clock fitting-time budgets within
each task to support a fair comparison with the four neural methods.
Physical-target validation error selects each task's configuration from its
candidate structures, visual dimensions and priors. The comparison evaluates
the forecasting performance of the resulting ViBR-WM model as a whole.
Sampling diagnostics check whether separate MCMC sampling
sequences (chains) give consistent posterior estimates. Effective sample
sizes estimate how many independent draws would provide the same precision
for posterior estimates as the correlated MCMC samples \citep{vehtari2021rank}.
Appendix~\ref{sec:representation} specifies these diagnostic criteria;
Table~\ref{tab:selected-configurations} lists the selected designs and priors.

\subsection{Metrics and uncertainty analysis}

Normalized root mean squared error (NRMSE) is computed by dividing prediction errors
by the corresponding target's training sample standard deviation, then taking the square
root of the mean squared normalized errors. For PointMaze and Wall, each fitted model's
NRMSE is computed over physical coordinates and forecast horizons within each test trajectory,
then averaged over trajectories and repetitions. For PhenoCam, NRMSE is computed across
test weeks separately for each camera site, Mead1 or Mead3, and forecast horizon,
then averaged equally over sites, horizons and repetitions. For SKIPP'D, NRMSE is computed
across forecast starting times separately at each horizon, then averaged over horizons
and repetitions.

We estimate uncertainty in mean NRMSE differences using 2,000 bootstrap resamples.
Each resample draws, with replacement, whole test trajectories for PointMaze and Wall
or time blocks for PhenoCam and SKIPP'D, together with prediction results from
each method's ten repetitions.
PhenoCam uses blocks of 13 consecutive weeks to preserve within-block temporal
dependence; a sensitivity analysis uses 26-week blocks to assess the effect of
block length. SKIPP'D groups predictions into 27 blocks by the week in which
forecasting starts, drawing 27 blocks with replacement per resample.
For each resample, we subtract each comparator's mean NRMSE from ViBR-WM's.
The 2.5th and 97.5th percentiles of these differences define a 95\% interval;
negative values indicate lower error for ViBR-WM.
These intervals apply separately to each comparison within the evaluated
environments, not jointly to all comparisons.
Section~\ref{sec:results} reports the overall intervals;
Appendices~\ref{sec:wall-native-final}--\ref{sec:scalar-native-final} give interval results
and resampling procedures.
Appendix~\ref{sec:native-probability}
evaluates predictive distributions, Appendix~\ref{sec:input-sensitivity} examines input and conversion
sensitivities, and Appendix~\ref{sec:synthetic} tests
variable selection and prediction of alternative futures from the same
observed history in controlled simulations.
\FloatBarrier

\section{Results}
\label{sec:results}

ViBR-WM achieves the lowest overall NRMSE among all five methods on all four
tasks, 16.19--50.37\% below the lowest-error comparator
(Figure~\ref{fig:native-overall}).

\subsection{Overall forecasting accuracy}

\begin{figure}[!t]
\centering
\includegraphics[width=\linewidth]{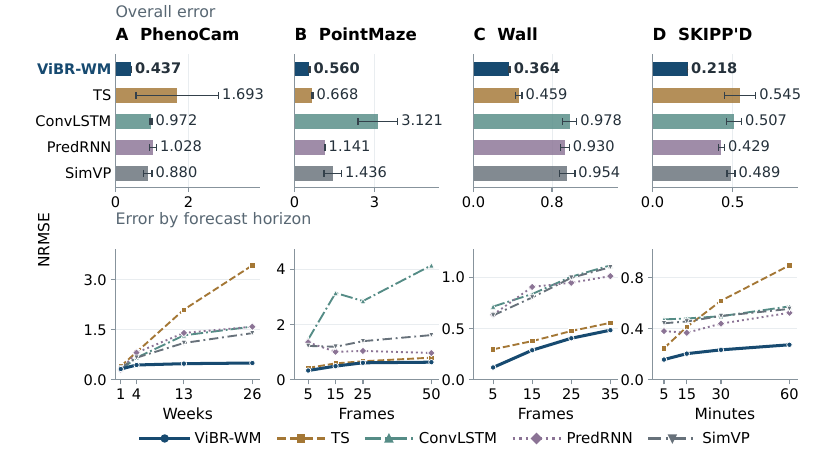}
\caption{Forecasting accuracy on four tasks (columns A--D).
Top: overall mean NRMSE across ten repetitions; error bars span one sample standard
deviation on either side of the mean.
Overall scores summarize four forecast horizons and, for PhenoCam, both camera sites.
Bottom: mean NRMSE by forecast horizon. PointMaze and Wall horizons give the number
of frames after forecasting starts at frame 10.
NRMSE normalizes errors by the corresponding training-target standard deviations.
Appendices~\ref{sec:wall-native-final}--\ref{sec:scalar-native-final}
report 95\% bootstrap intervals for differences between methods' mean errors.}
\label{fig:native-overall}
\label{fig:native-horizons}
\end{figure}

\noindent\textbf{PhenoCam.}
On the 2023 test data (Figure~\ref{fig:native-overall}A),
ViBR-WM's greenness NRMSE is 0.4367,
50.37\% below the lowest-error comparator, SimVP (0.8799), and 74.20\%
below Temporal Straightening.
Bootstrap resampling of 13-week blocks gives a 95\% interval
$[-0.7025,-0.2146]$ for ViBR-WM's mean NRMSE minus SimVP's.
The interval lies entirely below zero, supporting lower mean error for ViBR-WM.

\noindent\textbf{PointMaze.}
ViBR-WM's NRMSE is 0.5602 (Figure~\ref{fig:native-overall}B),
16.19\% below the best comparator, Temporal Straightening (0.6684),
with a 95\% bootstrap interval $[-0.1544,-0.0532]$ for ViBR-WM's mean
NRMSE minus TS's.

\noindent\textbf{Wall.}
For Wall position prediction
(Figure~\ref{fig:native-overall}C), ViBR-WM's NRMSE is 0.3640,
20.71\% below the best comparator, Temporal Straightening (0.4590),
with a 95\% interval $[-0.1554,-0.0350]$ for the NRMSE difference.

\noindent\textbf{SKIPP'D.}
On SKIPP'D's 2019 test data (Figure~\ref{fig:native-overall}D),
\methodname{} has mean NRMSE 0.2180: 49.20\% below the best comparator,
PredRNN (0.4290), with a 95\% interval $[-0.2384,-0.1871]$ for the NRMSE difference,
and 59.99\% below Temporal Straightening.

Across all four tasks, every 95\% bootstrap interval for ViBR-WM's overall mean
NRMSE minus each comparator's lies entirely below zero. For PhenoCam, this holds
with both 13-week and 26-week resampling blocks.

All selected ViBR-WM fits on PointMaze, Wall, PhenoCam and SKIPP'D pass the
MCMC checks for estimating forecasts. Coefficient checks also pass for every
PointMaze and Wall fit. For PhenoCam, coefficient and variable-selection checks
pass for all combinations of site, horizon and repetition; for SKIPP'D, they pass
for 39 of the 40 fits across ten repetitions and four horizons.
Appendices~\ref{sec:wall-native-final}--\ref{sec:scalar-native-final} report these checks,
scores for individual repetitions and error-difference intervals.

\subsection{Forecast accuracy by horizon and predictive uncertainty}

\begin{figure}[!t]
\centering
\includegraphics[width=0.78\linewidth]{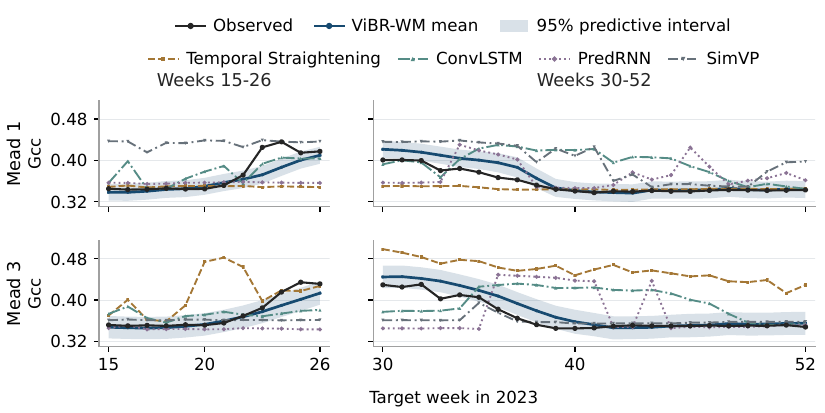}
\caption{Thirteen-week-ahead PhenoCam forecasts from repetition 1 at
the Mead 1 (top) and Mead 3 (bottom) camera sites. Columns show weeks 15--26 and 30--52
of 2023. Forecasts for weeks 27--29 are unavailable because the required input
data are incomplete.
Vertical axes show green chromatic coordinate (Gcc). Black curves show observed Gcc;
colored curves show predictions from the five methods. Blue shading denotes
ViBR-WM's 95\% predictive interval for each week.}
\label{fig:native-agriculture-examples}
\end{figure}

ViBR-WM has the lowest mean error at 4, 13 and 26 weeks ahead in PhenoCam
and at every evaluated horizon in PointMaze, Wall and SKIPP'D
(Figure~\ref{fig:native-horizons}).
Figure~\ref{fig:native-agriculture-examples} illustrates 13-week-ahead forecasts
at both PhenoCam camera sites.
The 95\% NRMSE-difference intervals against TS lie entirely below zero at
4, 13 and 26 weeks ahead in PhenoCam, 5, 15 and 50 frames ahead in PointMaze,
and 5 and 15 frames ahead in Wall.
For SKIPP'D, the 95\% interval against TS for the mean NRMSE difference averaged
over all four horizons is $[-0.3881,-0.2688]$, also entirely below zero.
Against ConvLSTM, PredRNN and SimVP,
the error-difference intervals lie entirely below zero at every evaluated horizon
in PointMaze and Wall, at 4, 13 and 26 weeks ahead in PhenoCam,
and for SKIPP'D's overall mean NRMSE.

\paragraph{Predictive uncertainty.}
ViBR-WM's 95\% predictive intervals contain, on average, 78.87\% of the observed target values in
PhenoCam, 93.66\% in PointMaze, 89.03\% in Wall and 97.43\% in SKIPP'D.
Appendix~\ref{sec:native-probability} evaluates these posterior predictive distributions
and compares variation across repeated fits of all five methods. This comparison
gives equal weight to each method's ten point predictions for a target; their spread
describes variation between fits, excluding future disturbances.
Continuous ranked probability scores (CRPS) and multivariate energy scores evaluate
the posterior predictive samples and repetition-based predictions against observations;
lower scores are better \citep{gneiting2007proper}.
For predictions across repeated fits, ViBR-WM has the lowest CRPS in PhenoCam
and SKIPP'D. On PointMaze and Wall, its CRPS and energy scores are lower than
ConvLSTM's, PredRNN's and SimVP's, while TS has the lowest scores.

\section{Discussion and conclusion}

ViBR-WM achieves the lowest overall NRMSE among all five methods on all four
tasks, 16.19--50.37\% below each task's best comparator. Repeated fitting and
resampling support these overall advantages.

Its architecture makes covariate effects interpretable through regression, with
distinct modules for trend, seasonal recurrence and stochastic cycles.
Visual compression and Bayesian selection reduce representation size and active
regression dimension. Posterior prediction incorporates parameter uncertainty and
future disturbances into recursive trajectories and direct target forecasts.
These results demonstrate that Bayesian regression can connect accurate world-model
forecasting with interpretable predictor effects and explicit uncertainty
quantification. Further evaluation can examine the structural modules in changing
environments and test whether posterior predictions help select actions for
reaching a goal.

\bibliography{references_arxiv}
\bibliographystyle{iclr2027_conference}

\clearpage
\input{ViBRWM_arxiv_appendix.tex}

\end{document}

%% file: figures/wall_forecasting_overview.tex
\begin{figure}[!t]
\centering
\includegraphics[width=\linewidth]{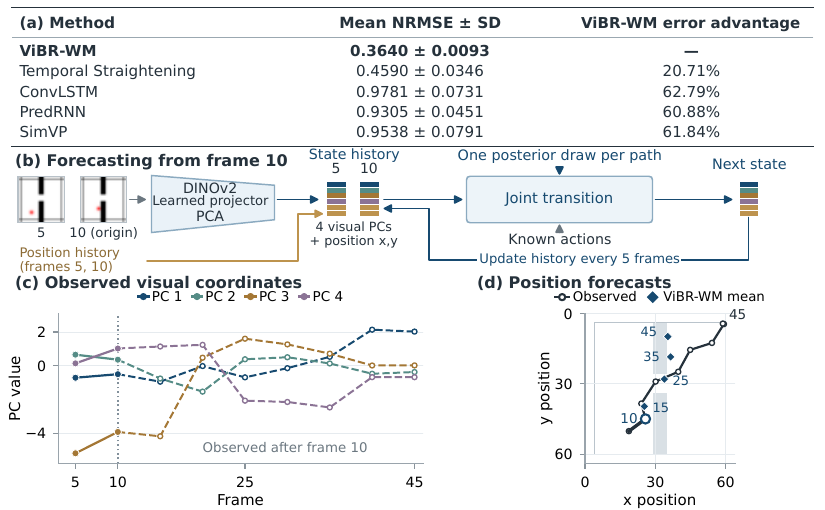}
\caption{Wall accuracy and forecasting.
(a) Overall NRMSE: mean $\pm$ sample standard deviation (SD) across ten fits,
with the lowest mean highlighted; percentages show how much lower ViBR-WM's
mean error is, expressed as a percentage of each comparator's mean error.
(b) Forecasting workflow for the test trajectory in (c--d).
DINOv2 extracts features from images at frames 5 and 10. These pass through
a projector trained with the TS objective. PCA then yields four visual
principal-component (PC) coordinates per frame. Each state-history column
contains these four visual coordinates and the observed
horizontal and vertical positions, $x$ and $y$, forming a six-dimensional state.
The joint-transition model uses the states at frames 5 and 10 and the known
actions to predict the state at frame 15. Each predicted state updates the history
for the next five-frame transition.
Each simulated trajectory uses one parameter set sampled from the posterior,
held fixed through all its transitions.
(c) Filled points show the PC values at frames 5 and 10 used in the history;
open points show the observed PC values after frame 10. (d) Observed positions and ViBR-WM
posterior means; labels identify frames.}
\label{fig:statistical-method}
\end{figure}

%% file: ViBRWM_arxiv_appendix.tex
\appendix
\setcounter{equation}{0}
\setcounter{table}{0}
\setcounter{figure}{0}
\renewcommand{\theequation}{S\arabic{equation}}
\renewcommand{\thetable}{S\arabic{table}}
\renewcommand{\thefigure}{S\arabic{figure}}
\renewcommand{\theHequation}{supp.\arabic{equation}}
\renewcommand{\theHtable}{supp.\arabic{table}}
\renewcommand{\theHfigure}{supp.\arabic{figure}}
\makeatletter
\setlength{\@fptop}{0pt}
\setlength{\@fpsep}{\floatsep}
\makeatother
\newcommand{\visionsubcontents}[1]{%
  \noindent\hspace*{1.5em}\hyperref[#1]{\ref*{#1}\quad\nameref*{#1}}\dotfill\pageref{#1}\par}
\section*{Appendix contents}
\noindent\hyperref[sec:world-model-inference]{A\quad Derivations for ViBR-WM}\dotfill\pageref{sec:world-model-inference}\par
\visionsubcontents{sec:visual-state-definition}
\visionsubcontents{sec:structural-modules}
\visionsubcontents{sec:likelihood-prior}
\visionsubcontents{sec:posterior-updates}
\visionsubcontents{sec:posterior-prediction-derivation}
\noindent\hyperref[sec:representation]{B\quad Data, visual reduction and common evaluation}\dotfill\pageref{sec:representation}\par
\noindent\hyperref[sec:wall-native-final]{C\quad Wall: specification and results}\dotfill\pageref{sec:wall-native-final}\par
\noindent\hyperref[sec:pointmaze-native-final]{D\quad PointMaze: specification and results}\dotfill\pageref{sec:pointmaze-native-final}\par
\noindent\hyperref[sec:scalar-native-final]{E\quad PhenoCam and SKIPP'D: specifications and results}\dotfill\pageref{sec:scalar-native-final}\par
\visionsubcontents{sec:phenocam-specification}
\visionsubcontents{sec:sky-native-returned}
\noindent\hyperref[sec:native-probability]{F\quad Predictive distributions and repeat-fit dispersion}\dotfill\pageref{sec:native-probability}\par
\noindent\hyperref[sec:input-sensitivity]{G\quad Input and conversion sensitivity}\dotfill\pageref{sec:input-sensitivity}\par
\noindent\hyperref[sec:synthetic]{H\quad Controlled diagnostics}\dotfill\pageref{sec:synthetic}\par
\visionsubcontents{sec:known-support-dynamics}
\visionsubcontents{sec:multiple-future-prediction}
\noindent\hyperref[sec:native-resources]{I\quad Resources and reproducibility}\dotfill\pageref{sec:native-resources}\par

\clearpage
\section{Derivations for ViBR-WM}
\label{sec:world-model-inference}

\subsection{Visual observations and predictive states}
\label{sec:visual-state-definition}

For PointMaze and Wall, the Visual Bayesian Regression World Model
(ViBR-WM) combines visual features with observed physical coordinates.
Let $o_t$ denote an image with red, green and blue (RGB) channels. Its frozen
DINOv2 patch field is $e_t\in\mathbb{R}^{196\times384}$. A learned projector,
trained with the Temporal Straightening (TS) objective, produces
$z_t^v=\mathcal{P}_\phi(e_t)
\in\mathbb{R}^{196\times8}$ \citep{wang2026straightening}.
Each row represents one spatial patch and contains eight feature channels.

\paragraph{From patch features to visual coordinates.}
The map $\mathcal R_{\mathrm{train}}$ reduces the patch field to $q$ visual
coordinates through channel principal component
analysis (PCA), spatial
concatenation, a second PCA and coordinate standardization.
Let $\mu_c\in\mathbb R^8$ be the mean over all training patches and
$C\in\mathbb R^{8\times c}$ contain the leading channel-PCA directions.
Writing $\mathbf1_{196}$ for a vector of 196 ones, the concatenated field is
\[
 w_t=\operatorname{vec}\!\left(
       [(z_t^v-\mathbf1_{196}\mu_c^\top)C]^\top\right)
       \in\mathbb R^{196c}.
\]
Here $\operatorname{vec}$ stacks matrix columns, so this expression lists
the $c$ retained channels of the first patch, then the second, and so on.
PointMaze and Wall use $c=7$, giving $196c=1372$ entries per frame.
The projector's final layer normalization constrains the eight channels
to an affine subspace of dimension at most seven.

Let $\mu_w\in\mathbb R^{196c}$ be the training-frame mean of $w_t$.
Center the training differences $w_{t+\Delta}-w_t$, computed within each
video, and collect them as rows of a matrix. Its leading $q$ right singular
vectors form the columns of $V\in\mathbb R^{196c\times q}$.
With $\Delta=5$ for PointMaze and Wall, define
\[
 b_t=V^\top(w_t-\mu_w)\in\mathbb R^q,\qquad
 \mathcal R_{\mathrm{train}}(z_t^v)=D_v^{-1}(b_t-\mu_b)
 =\widetilde z_t^v.
\]
The vector $\mu_b$ is the training mean of $b_t$, and $D_v$ is diagonal
with the sample standard deviations of $b_{t+\Delta}-b_t$.
Thus differences determine the projection directions and scales, while each
frame's feature levels supply its coordinates. The means, directions and
scales are fitted once and held fixed during forecasting.

\paragraph{Why use increments for the second PCA?}
Let $D_\Delta$ collect the centered training differences
$w_{t+\Delta}-w_t$ as rows. The chosen basis solves
\[
 \max_{V^\top V=I_q}\ \|D_\Delta V\|_F^2,
 \qquad\text{equivalently}\qquad
 \min_{V^\top V=I_q}\ \|D_\Delta-D_\Delta VV^\top\|_F^2.
\]
Here $\|\cdot\|_F^2$ sums squared matrix entries. The criterion retains
the largest variation in changes over the modeled transition interval;
level PCA would instead retain variation in the frame features themselves.
Projecting levels with this basis gives coordinates aligned with the dominant
training changes. The dimension comparisons below vary $q$ while using the same
variance-based rule to select the projection directions.

\paragraph{Physical coordinates and joint state.}
For $p_t\in\mathbb R^d$, let $\mu_p\in\mathbb R^d$ contain coordinatewise
training means and let $D_p\in\mathbb R^{d\times d}$ be diagonal with their
sample standard deviations over all training frames. Then
\[
 \bar p_t=D_p^{-1}(p_t-\mu_p),\qquad
 s_t=[(\widetilde z_t^v)^\top,\bar p_t^\top]^\top
 \in\mathbb R^m,\quad m=q+d.
\]
The bar denotes standardization, not temporal averaging. The inverse map
$p_t=\mu_p+D_p\bar p_t$ returns forecasts to physical units.

\paragraph{Histories, inputs and responses.}
PointMaze and Wall actions are $a_t\in\mathbb R^2$, with the five-frame block
$A_t=[a_t^\top,\ldots,a_{t+4}^\top]^\top\in\mathbb R^{10}$.
The additive design uses $A_t,A_{t-10},A_{t-5}$, giving thirty action values.
Thus $K=3$ action blocks are retained independently of the $L$ state vectors
in the main-text history definition. We reserve $P$ for the number of regressors
\emph{per response}. These differ from the visual dimension $q$ and the
total number $mP$ of response-specific coefficients.

Let $\mathcal H_t$ contain the state history and known action blocks.
The map $\chi$ concatenates their values and appends any state--action
products and position-dependent regional terms specified by the
validation-chosen design to form $x_t=\chi(\mathcal H_t)\in\mathbb R^P$.
The regional terms let velocity and action effects vary smoothly with position;
their formulas are given in Appendix~\ref{sec:pointmaze-native-final}.
Let $\mu_\theta(x_t)$ be the conditional mean state increment and
$f_\theta$ the corresponding next-state mean. The notation
$o_t,a_t,e_t,z_t^v,\mathcal P_\phi$ and $f_\theta$ follows the observation,
action, representation and prediction roles in \citet{wang2026straightening}.
Here $s_t$ augments the reduced visual representation with physical coordinates.
The fitted projector and reduction are fixed during Bayesian inference;
$\theta=(B,\Sigma,\gamma)$ collects regression coefficients, disturbance
covariance and inclusion indicators, defined below.
The model for PointMaze and Wall is
\begin{align}
 f_\theta(\mathcal H_t)&=s_t+\mu_\theta(x_t),\\
 s_{t+5}\mid\mathcal H_t,\theta
 &\sim\mathcal N_m\!\left(f_\theta(\mathcal H_t),\Sigma\right).
 \label{eq:wm-transition}
\end{align}
\begin{samepage}
For PhenoCam and SKIPP'D, $\mathcal I_{t,h}$ contains image and numerical observations
available at the forecast origin and calendar covariates known for the target
time. The direct model is
$u_{t+h}\mid\mathcal I_{t,h},\theta_h\sim
\mathcal N(\mu_{\theta_h}(\chi_h(\mathcal I_{t,h})),\sigma_h^2)$, where
$u_{t+h}=T_h(y_{t+h})$. For PhenoCam green chromatic coordinate (Gcc),
$T_h(g)=(g-\mu_{g,h})/d_{g,h}$: the mean and sample standard deviation
come from that site's horizon-specific training-response rows. Historical
Gcc regressors use the same transformation; original-unit predictions are
$g=\mu_{g,h}+d_{g,h}u$. SKIPP'D uses $T_h(p)=p$ for raw photovoltaic (PV) power.
Each site/horizon regression has its own scalar predictive distribution.
\par
\end{samepage}

\subsection{Structural modules and their joint prediction}
\label{sec:structural-modules}

Trend, seasonal and cycle states follow distinct transition rules and
combine with regression in the following observation equation.
For response $j$ at sequence step $t$,
\begin{equation}
 r_{tj}=\bar r_j+\ell_{tj}+\zeta_{tj}+\psi_{tj}
       +\beta_j^\top\check x_{tj}+\epsilon_{tj},
 \qquad\epsilon_t\sim\mathcal N_m(0,\Sigma).
\end{equation}
Here $r_{tj}$ is the modeled response at time $t$, $\bar r_j$ is its fixed
training center, and $\beta_j^\top\check x_{tj}$ is the regression contribution
from the centered predictors $\check x_{tj}$ defined below.
The three states $\ell_{tj}$, $\zeta_{tj}$ and $\psi_{tj}$ denote trend,
seasonal and cycle contributions. The observation disturbance is $\epsilon_{tj}$.
Within this subsection, $t$ counts successive observations in one regularly
sampled sequence, rather than individual video frames; inactive structural
modules are zero.
\paragraph{Trend: evolving level and growth.}
The trend module tracks a changing baseline and its rate of change through
a level and a potentially damped slope:
\begin{align}
 \ell_{t+1,j}&=\ell_{tj}+d_{tj}+\eta^\ell_{tj}, &
 d_{t+1,j}&=\rho_jd_{tj}+\eta^d_{tj}.
\end{align}
Here $d_{tj}$ is the slope, and $\eta^\ell_{tj}$ and $\eta^d_{tj}$ are
random disturbances to level and slope. The slope is undamped for $\rho_j=1$ and damped toward zero for
$0<\rho_j<1$. A local-level specification omits the slope state.
\paragraph{Seasonality: recurring patterns that can evolve.}
For an integer seasonal period $S_j\ge2$,
\begin{equation}
 \zeta_{t+1,j}=-\sum_{k=0}^{S_j-2}\zeta_{t-k,j}+\eta^\zeta_{tj}.
\end{equation}
Equivalently, $\sum_{k=0}^{S_j-1}\zeta_{t+1-k,j}=\eta^\zeta_{tj}$.
Thus a complete seasonal period sums to zero without disturbances;
the random term $\eta^\zeta_{tj}$ permits the seasonal pattern to evolve.
\paragraph{Cycle: oscillation and persistence after disturbances.}
The cycle uses a two-dimensional rotating state. Its first coordinate
$\psi_{tj}$ enters the observation equation; the auxiliary coordinate
$\psi^*_{tj}$ allows the pair to rotate:
\begin{equation}
 \begin{pmatrix}\psi_{t+1,j}\\\psi^*_{t+1,j}\end{pmatrix}
 =\varrho_j
 \begin{pmatrix}\cos\lambda_j&\sin\lambda_j\\
                -\sin\lambda_j&\cos\lambda_j\end{pmatrix}
 \begin{pmatrix}\psi_{tj}\\\psi^*_{tj}\end{pmatrix}
 +\begin{pmatrix}\eta^\psi_{tj}\\\eta^{\psi^*}_{tj}\end{pmatrix}.
\end{equation}
The two $\eta$ terms perturb the cycle coordinates.
The specified frequency $0<\lambda_j<\pi$ gives nominal period
$2\pi/\lambda_j$, with $0<\varrho_j\le1$ controlling damping.
An isolated state perturbation evolves after $k$ steps by
$\varrho_j^k R_j^k$, where $R_j$ is the displayed orthogonal rotation;
its norm therefore decays as $\varrho_j^k$ when $\varrho_j<1$.
Thus a disturbance can have a persistent oscillating effect, whose decay rate
is governed by $\varrho_j$. The frequency controls recurrence separately
from the seasonal period.
\paragraph{Joint inference and prediction.}
State disturbances are zero-mean Gaussian with diagonal covariance. The
two cycle disturbance variances within each response can be estimated
separately or constrained to be equal. Enabled
states and regression coefficients are inferred jointly. Posterior forecasts
sum the structural and regression contributions and draw future state and
observation disturbances.

\subsection{Likelihood and prior}
\label{sec:likelihood-prior}

The following derivation concerns the regression module.
Let $i=1,\ldots,N$ index training windows and $t_i$ their forecast origins.
For PointMaze and Wall, $x_i=\chi(\mathcal H_{t_i})$ and
$r_i=s_{t_i+5}-s_{t_i}\in\mathbb{R}^m$ are the input and response.
For a direct forecast, they are $x_i=\chi_h(\mathcal I_{t_i,h})$ and
the transformed scalar response $r_i=u_{t_i+h}$.
Each scalar fit uses the following equations with $m=1$ and $\Sigma=\sigma_h^2$.
For response coordinate $j=1,\ldots,m$,
\begin{align}
 \check x_{ij}&=D_x^{-1}(x_i-\mu_x)-c_{x,j},\\
 r_{ij}&=\bar r_j+\beta_j^\top\check x_{ij}+\epsilon_{ij},
 \qquad\epsilon_i\sim\mathcal N_m(0,\Sigma).
 \label{eq:wm-centered-regression}
\end{align}
Here $r_{ij}$ is coordinate $j$ of response $r_i$, and
$\beta_j\in\mathbb R^P$ assigns one coefficient to each predictor for that coordinate.
The vector $\mu_x\in\mathbb R^P$ and diagonal matrix
$D_x\in\mathbb R^{P\times P}$ give the predictor means and scales.
After this transformation, $c_{x,j}\in\mathbb R^P$ averages the transformed
inputs across training windows, and $\bar r_j\in\mathbb R$ averages their responses $r_{ij}$.
The additional center records the regression fit's internal centering.
With the same complete, standardized inputs for every response,
all $c_{x,j}$ coincide and are zero apart from numerical rounding.
For direct forecasts, the main text denotes these centers by $c_{x,h}$ and $\bar u_h$.
These centers and scales are fixed; the conditional mean for a new input is
$\mu_{\theta,j}(x)=\bar r_j+\beta_j^\top\{D_x^{-1}(x-\mu_x)-c_{x,j}\}$.
When no separate design standardization is applied, as in the scalar
PhenoCam and SKIPP'D regressions, $\mu_x=0$ and $D_x=I$; internal centering
still applies to the preconstructed features.
The response mean $\bar r_j$ stays fixed and coefficients are time-invariant.

\paragraph{Stacking the response equations.}
For $N$ training windows, let $X_j\in\mathbb{R}^{N\times P}$ have rows
$\check x_{ij}^\top$ and let $R\in\mathbb R^{N\times m}$ have rows $r_i^\top$.
Write $\bar r=(\bar r_1,\ldots,\bar r_m)^\top\in\mathbb R^m$ and
$\mathbf1_N\in\mathbb R^N$ for a vector of ones. Define
\[
 \mathcal X=\operatorname{blockdiag}(X_1,\ldots,X_m)
 \in\mathbb R^{Nm\times mP},\qquad
 y=\operatorname{vec}(R-\mathbf1_N\bar r^\top)\in\mathbb R^{Nm}.
\]
The block-diagonal matrix places $X_1,\ldots,X_m$ along its diagonal
and zeros elsewhere; column-stacking orders $y$ by response, with all $N$
values of response 1 first. Collect coefficients as rows of
$B=[\beta_1,\ldots,\beta_m]^\top\in\mathbb R^{m\times P}$;
entry $\beta_{jk}$ weights predictor $k$ for response $j$.
Then $\beta=\operatorname{vec}(B^\top)\in\mathbb{R}^{mP}$ lists all
$P$ coefficients for response 1, then response 2, and so on,
matching the columns of $\mathcal X$.
For predictor $k$ and response $j$, $\gamma_{jk}\in\{0,1\}$ indicates
inclusion. Collect these indicators in the same order as $\beta$, and let
$k_\gamma=\sum_{j=1}^m\sum_{k=1}^P\gamma_{jk}$.
Then $\mathcal X_\gamma\in\mathbb R^{Nm\times k_\gamma}$ retains the included
columns and $\beta_\gamma\in\mathbb R^{k_\gamma}$ their coefficients.

The stacked disturbance covariance is
$\Omega=\Sigma\otimes I_N\in\mathbb R^{Nm\times Nm}$, where $I_N$ is
the identity matrix and $\otimes$ denotes the Kronecker product.
Its response block $(j,\ell)$ is $\Sigma_{j\ell}I_N$: disturbances can
correlate across responses in the same row but are modeled as independent
between rows. Stacking the response equations gives
$y=\mathcal X_\gamma\beta_\gamma+\varepsilon$ with
$\varepsilon\sim\mathcal N_{Nm}(0,\Omega)$.
The Kronecker identities
$|\Omega|=|\Sigma|^N$ and $\Omega^{-1}=\Sigma^{-1}\otimes I_N$
then give the Gaussian working likelihood, conditional on the inclusion vector
$\gamma$:
\begin{equation}
 p(y\mid\beta_\gamma,\Sigma,\gamma)
 \propto |\Sigma|^{-N/2}
 \exp\!\left[-\tfrac12
 (y-\mathcal X_\gamma\beta_\gamma)^\top\Omega^{-1}
 (y-\mathcal X_\gamma\beta_\gamma)\right].
 \label{eq:wm-likelihood}
\end{equation}
For a square matrix, subscript $\gamma$ selects its included rows and columns.
Excluded coefficients are set to zero. For the selected fits,
the centered response-specific design blocks have full column rank.
The likelihood treats training rows as conditionally independent.
Overlapping histories and target intervals may induce residual dependence
absent from $\Omega$. Posterior covariance and inclusion probabilities
therefore depend on this working approximation. Inference also conditions
on the fitted representation, transformations, feature map and prior
hyperparameters; their estimation and selection uncertainty is not propagated.

The implemented prior has independent inclusion indicators:
\begin{align}
 \gamma_{jk}&\sim\operatorname{Bernoulli}(\pi_{jk}),
 &\beta_\gamma\mid\gamma&\sim
 \mathcal N(0,\Lambda_{0,\gamma}^{-1}),\\
 \Lambda_0&=\frac{\kappa}{N}\mathcal X^\top\mathcal X,
 &\Sigma&\sim\operatorname{IW}_m(\nu_0,V_0),\\
 V_0&=(\nu_0-m-1)(1-R^2)S_r,
 &\nu_0&=m+2,
 \label{eq:wm-prior}
\end{align}
Here $\pi_{jk}$ is the prior inclusion probability.
The precision $\Lambda_{0,\gamma}$ is the inverse coefficient covariance;
it selects included rows and columns from the $mP\times mP$ matrix $\Lambda_0$.
Full column rank of the design makes these precision matrices invertible.
The notation $\operatorname{IW}_m(\nu_0,V_0)$ denotes the inverse-Wishart
distribution on covariance matrices, with degrees of freedom $\nu_0$ and scale $V_0$.
The matrix $S_r\in\mathbb R^{m\times m}$ is the sample covariance of training responses;
$R^2$ is a specified prior hyperparameter, not a goodness-of-fit score measured
after fitting. The constant $\kappa>0$ controls prior precision.
Larger $\kappa$ tightens the prior
around zero. The covariance prior has mean
$\mathbb E[\Sigma]=V_0/(\nu_0-m-1)=(1-R^2)S_r$,
so $1-R^2$ specifies the fraction of response covariance assigned to
unexplained disturbances a priori. The coefficient
prior is independent of $\Sigma$, conditional on $\gamma$ and the fixed
training quantities. We use $S_r\succ0$ and $0\le R^2<1$, so $V_0$ is
positive definite. The inverse-Wishart convention is
$p(\Sigma)\propto|\Sigma|^{-(\nu_0+m+1)/2}
\exp[-\operatorname{tr}(V_0\Sigma^{-1})/2]$.
PointMaze and Wall use fixed $\pi_{jk}=1$ for every term in the selected design.
Mixed-inclusion scalar regressions fix the inclusion probabilities of
numerical-history and calendar regressors at one and allow visual terms
to be selected. Inclusion is learned for the selectable visual terms;
their posterior inclusion probabilities quantify predictive relevance
conditional on the selected design.
Write $\theta=(B,\Sigma,\gamma)$ for the general regression parameters and
$\theta_h$ for the scalar counterparts. PointMaze and Wall fix $\gamma$ to one,
so their free parameters reduce to $(B,\Sigma)$.

\paragraph{Interpretable effects and active dimension.}
For a fixed input, $\check x_{ijk}$ is entry $k$ of the centered vector
$\check x_{ij}$. In each posterior draw, $\beta_{jk}\check x_{ijk}$ is
predictor $k$'s contribution to response $j$'s conditional mean.
Summaries across draws quantify its uncertainty. State--action products and
position-weighted regional terms have this interpretation on their constructed feature scale; changing one
raw variable can change several such terms.
For training data $\mathcal D$, each response has an active dimension
$P_{\mathrm{active},j}=\sum_{k=1}^P\gamma_{jk}$, with posterior expectation
\[
 \mathbb E[P_{\mathrm{active},j}\mid\mathcal D]
 =\sum_{k=1}^P\Pr(\gamma_{jk}=1\mid\mathcal D).
\]
This makes regression dimension a posterior quantity when inclusion is
learned. Prediction averages over the sampled subsets and coefficients,
preserving uncertainty about which variables contribute.

\subsection{Derivation of the posterior updates}
\label{sec:posterior-updates}

The calculations condition on the design, training transforms and prior
hyperparameters; these fixed quantities are suppressed in the notation.
Here $y$ is the centered response vector defined above.
Bayes' rule gives
$p(\beta_\gamma,\Sigma,\gamma\mid y)\propto
p(y\mid\beta_\gamma,\Sigma,\gamma)
p(\beta_\gamma\mid\gamma)p(\Sigma)p(\gamma)$.

\paragraph{Regression coefficients: completing the square.}
For fixed $\Sigma$ and $\gamma$, define the posterior precision
$Q_\gamma\in\mathbb R^{k_\gamma\times k_\gamma}$ and information vector
$b_\gamma\in\mathbb R^{k_\gamma}$:
\begin{align}
 Q_\gamma&=\Lambda_{0,\gamma}
       +\mathcal X_\gamma^\top\Omega^{-1}\mathcal X_\gamma,
 &b_\gamma&=\mathcal X_\gamma^\top\Omega^{-1}y.
\end{align}
The precision combines the prior precision with information from the data;
its inverse will be the conditional posterior covariance.
Multiplying the likelihood by the zero-mean Gaussian prior adds their
quadratic exponents. Expanding and completing the square yields
\begin{align}
 &(y-\mathcal X_\gamma\beta_\gamma)^\top\Omega^{-1}
   (y-\mathcal X_\gamma\beta_\gamma)
   +\beta_\gamma^\top\Lambda_{0,\gamma}\beta_\gamma \notag\\
 &\quad=y^\top\Omega^{-1}y-2\beta_\gamma^\top b_\gamma
        +\beta_\gamma^\top Q_\gamma\beta_\gamma \notag\\
 &\quad=(\beta_\gamma-m_\gamma)^\top Q_\gamma
          (\beta_\gamma-m_\gamma)
        +y^\top\Omega^{-1}y-b_\gamma^\top Q_\gamma^{-1}b_\gamma,
 \label{eq:wm-square-completion}
\end{align}
where $m_\gamma=Q_\gamma^{-1}b_\gamma$.
Here $m_\gamma\in\mathbb R^{k_\gamma}$ is the conditional posterior mean.
The last two terms do not involve $\beta_\gamma$. Normalizing the remaining
Gaussian kernel therefore gives
\begin{equation}
 \beta_\gamma\mid\Sigma,\gamma,y
 \sim\mathcal N(m_\gamma,Q_\gamma^{-1}).
 \label{eq:wm-coef-conditional}
\end{equation}
Cross-response covariance enters $Q_\gamma$ through $\Omega^{-1}$.
Before restricting to included coefficients, its likelihood information
block $(j,\ell)$ is $(\Sigma^{-1})_{j\ell}X_j^\top X_\ell$.
Thus each fit for PointMaze or Wall couples coefficient inference across responses.

\paragraph{Disturbance covariance: collecting the residual terms.}
Let $\widehat R_\beta\in\mathbb R^{N\times m}$ have column $j$ equal to
$X_j\beta_j$, with excluded coefficients set to zero. The residual matrix is
$E=R-\mathbf1_N\bar r^\top-\widehat R_\beta$, so the stacked residual is
$y-\mathcal X_\gamma\beta_\gamma=\operatorname{vec}(E)$.
The likelihood quadratic can be written as
\begin{equation}
 \operatorname{vec}(E)^\top(\Sigma^{-1}\otimes I_N)\operatorname{vec}(E)
 =\operatorname{tr}(E^\top E\Sigma^{-1}).
\label{eq:wm-residual-trace}
\end{equation}
Here $\operatorname{tr}$ sums diagonal entries. Entry $(j,\ell)$ of
$E^\top E$ is $\sum_{i=1}^N E_{ij}E_{i\ell}$, collecting residual
cross-products for responses $j$ and $\ell$.
Multiplying the likelihood by the inverse-Wishart prior combines the
determinant powers and adds the scale matrices:
\begin{equation}
 p(\Sigma\mid\beta,\gamma,y)\propto
 |\Sigma|^{-(\nu_0+N+m+1)/2}
 \exp\!\left[-\tfrac12
   \operatorname{tr}\{(V_0+E^\top E)\Sigma^{-1}\}\right].
 \label{eq:wm-cov-kernel}
\end{equation}
This is the inverse-Wishart density with update
\begin{equation}
 \Sigma\mid\beta,\gamma,y\sim
 \operatorname{IW}_m(\nu_0+N,V_0+E^\top E).
 \label{eq:wm-cov-conditional}
\end{equation}
The coefficient prior contributes no determinant power of $\Sigma$
because it is independent of $\Sigma$. The degrees of freedom increase
by $N$, with no additional coefficient-count term.

\paragraph{Variable inclusion: integrating out the coefficients.}
For the $k_\gamma$ included coefficients, the normalized prior contributes
$(2\pi)^{-k_\gamma/2}|\Lambda_{0,\gamma}|^{1/2}$.
Using Equation~(\ref{eq:wm-square-completion}), the required integral is
\begin{equation}
 \int_{\mathbb R^{k_\gamma}}
 \exp\!\left[-\tfrac12(\beta_\gamma-m_\gamma)^\top
                     Q_\gamma(\beta_\gamma-m_\gamma)\right]d\beta_\gamma
 =(2\pi)^{k_\gamma/2}|Q_\gamma|^{-1/2}.
 \label{eq:wm-gaussian-integral}
\end{equation}
The powers of $2\pi$ cancel. Conditional on $\Sigma$, the factors
$|\Sigma|^{-N/2}$ and $\exp(-y^\top\Omega^{-1}y/2)$ are common to all
inclusion patterns. Dropping only these common factors gives
\begin{equation}
 \omega(\gamma;\Sigma,y)=
 p(\gamma)\frac{|\Lambda_{0,\gamma}|^{1/2}}{|Q_\gamma|^{1/2}}
 \exp\!\left(\tfrac12 b_\gamma^\top Q_\gamma^{-1}b_\gamma\right).
 \label{eq:wm-inclusion}
\end{equation}
The weight $\omega$ is proportional to the conditional posterior probability
of a predictor subset. Its prior factor $p(\gamma)$ multiplies $\pi_{jk}$
for included terms and $1-\pi_{jk}$ for excluded terms.
Integrating over coefficients accounts for their possible values when
comparing subsets. Normalizing over all subsets allowed by the fixed indicators gives
$$p(\gamma\mid\Sigma,y)=\omega(\gamma;\Sigma,y)/\sum_{\gamma'}\omega(\gamma';\Sigma,y).$$
For a selectable indicator $\gamma_{jk}$, keep all other indicators fixed
and let $\omega_1,\omega_0$ be the two weights with that indicator included or excluded.
Writing $\gamma_{-(jk)}$ for all the other indicators, normalizing these
two alternatives gives the update
\begin{equation}
 \Pr(\gamma_{jk}=1\mid\gamma_{-(jk)},\Sigma,y)=\frac{\omega_1}{\omega_0+\omega_1}.
 \label{eq:wm-single-inclusion}
\end{equation}
Indicators with $\pi_{jk}=0$ or $1$ stay fixed. Empty selected sets have
determinant one and quadratic form zero. A sampling sweep updates the
selectable indicators conditional on $\Sigma$, draws the coefficients
conditional on the new indicators and current $\Sigma$, and then draws $\Sigma$ conditional
on those coefficients. In the PointMaze and Wall fits all indicators are fixed,
so only the coefficient and covariance updates are needed.

\paragraph{Scalar specialization.}
For a direct target, set $m=1$, $\Sigma=\sigma_h^2$,
$X_\gamma=(X_1)_\gamma=\mathcal X_\gamma$ and
$e=y-X_\gamma\beta_\gamma$. The same calculation gives
\begin{align}
 Q_\gamma&=\Lambda_{0,\gamma}+\frac{X_\gamma^\top X_\gamma}{\sigma_h^2},
 &b_\gamma&=\frac{X_\gamma^\top y}{\sigma_h^2},\\
 \sigma_h^2\mid\beta,\gamma,y
 &\sim\operatorname{IG}\!\left(\frac{\nu_0+N}{2},
                              \frac{V_0+e^\top e}{2}\right).
 \label{eq:wm-scalar-conditional}
\end{align}
Here $\operatorname{IG}(a,b)$ denotes the inverse-gamma distribution, with density proportional to
$v^{-a-1}\exp(-b/v)$ for $v>0$; $\nu_0=3$ in the scalar fits.
The Gaussian coefficient prior remains independent of $\sigma_h^2$.
These are conditional posterior updates; prediction integrates their
joint posterior numerically.

\subsection{Derivation of posterior prediction and uncertainty}
\label{sec:posterior-prediction-derivation}

Let $\mathcal H_t$ denote the observed origin history and $\mathcal A$
the known action schedule; recall that $\mathcal D$ denotes the training data.

\paragraph{One-step predictive distribution.}
For a fixed forecast input $x_*$, the centered input for response $j$ is
$\check x_j(x_*)=D_x^{-1}(x_*-\mu_x)-c_{x,j}$.
Define the response-specific design
$$\mathcal X_*=\operatorname{blockdiag}
(\check x_1(x_*)^\top,\ldots,\check x_m(x_*)^\top)
\in\mathbb R^{m\times mP}.$$
The vector $r_*\in\mathbb R^m$ is the next response to predict at that input.
Then $r_*=\bar r+\mathcal X_{*,\gamma}\beta_\gamma+\epsilon_*$,
where $\epsilon_*\sim\mathcal N_m(0,\Sigma)$ is independent of the
coefficient draw conditional on $\Sigma,\gamma,\mathcal D$.
An affine transformation of the Gaussian coefficient conditional in
Equation~(\ref{eq:wm-coef-conditional}), followed by adding this independent
disturbance, gives
\begin{equation}
 r_*\mid\Sigma,\gamma,\mathcal D,x_*
 \sim\mathcal N_m\!\left(
 \bar r+\mathcal X_{*,\gamma}m_\gamma,
 \Sigma+\mathcal X_{*,\gamma}Q_\gamma^{-1}\mathcal X_{*,\gamma}^\top
 \right).
 \label{eq:wm-one-step-predictive}
\end{equation}
Predictive covariance includes both future
disturbance variance and coefficient uncertainty. Averaging this density
over $p(\Sigma,\gamma\mid\mathcal D)$ also incorporates covariance and
inclusion uncertainty. For PointMaze and Wall, translating $r_*$ by $s_t$ gives
the next-state distribution.

\paragraph{Recursive trajectories.}
Conditional on a parameter draw, the chain rule factors the density of
an $H$-transition path into the successive state-transition densities.
Integrating that conditional density over the posterior gives
\begin{equation}
 p(\{s_{t+5\ell}\}_{\ell=1}^H\mid\mathcal H_t,\mathcal A,\mathcal D)
 =\int\prod_{\ell=0}^{H-1}
 p_\theta(s_{t+5(\ell+1)}\mid\mathcal H_{t+5\ell}^{\mathrm{roll}},\mathcal A)
 \,p(\theta\mid\mathcal D)\,d\theta.
 \label{eq:wm-predictive}
\end{equation}
The index $\ell$ counts five-frame transitions.
The history $\mathcal H_{t+5\ell}^{\mathrm{roll}}$ contains the prescribed
number of recent observed or simulated states and known action blocks,
starting from $\mathcal H_t^{\mathrm{roll}}=\mathcal H_t$.
Here $\mathcal A$ supplies the action schedule and $\mathcal D$ is the training data.
After each transition, append that path's sampled
state and construct its next input from the prescribed history and known actions.
Each path retains one parameter draw and samples a new disturbance at each step.
Physical forecasts apply the inverse training standardization to the physical block
of $s$. Reusing the parameter draw along a path preserves the dependence
induced by parameter uncertainty. For example, consider the linear transition
$s_{t+5}=As_t+\epsilon_t$, where $A$ is a transition matrix and the disturbances
have mean zero. The two-step conditional mean is $A^2s_t$. Its posterior average is
$\mathbb E[A^2\mid\mathcal D]s_t$, generally different from
$\mathbb E[A\mid\mathcal D]^2s_t$. Recursively applying posterior-mean
parameters therefore need not give the posterior predictive mean.
For PointMaze, the position-dependent regional weights additionally require evaluating the
design on each sampled state history.

\paragraph{Direct forecasts and total predictive variance.}
Direct regressions instead integrate
$p_{\theta_h}(u_{t+h}\mid\mathcal I_{t,h})$ against
$p(\theta_h\mid\mathcal D)$ without feeding back earlier predicted targets.
For $x_*=\chi_h(\mathcal I_{t,h})$, the laws of total expectation and variance give
\begin{align}
 \mathbb E[u_{t+h}\mid\mathcal I_{t,h},\mathcal D]
 &=\mathbb E_{\theta_h\mid\mathcal D}[\mu_{\theta_h}(x_*)],\\
 \operatorname{Var}(u_{t+h}\mid\mathcal I_{t,h},\mathcal D)
 &=\mathbb E_{\theta_h\mid\mathcal D}[\sigma_h^2]
   +\operatorname{Var}_{\theta_h\mid\mathcal D}[\mu_{\theta_h}(x_*)].
 \label{eq:wm-total-predictive-variance}
\end{align}
The first term averages future disturbance variance over the posterior;
the second measures variation in the regression mean across coefficient and
inclusion draws. For the inverse target transform $y=\mu_{y,h}+d_{y,h}u$,
use $(\mu_{y,h},d_{y,h})=(\mu_{g,h},d_{g,h})$ for Gcc and $(0,1)$ for SKIPP'D.
The predictive mean becomes
$\mu_{y,h}+d_{y,h}\mathbb E[u\mid\mathcal I_{t,h},\mathcal D]$ and variance becomes
$d_{y,h}^2\operatorname{Var}(u\mid\mathcal I_{t,h},\mathcal D)$.
Each target/horizon fit defines its own marginal predictive distribution;
these separate fits do not specify cross-horizon dependence or a joint
probability for an entire future path.
PhenoCam point estimates average conditional means; SKIPP'D point estimates
average predictive samples including observation noise.
Both estimate the posterior predictive mean; sampling observation noise
adds finite-sample simulation variability.
PointMaze and Wall point estimates average sampled paths. Section~\ref{sec:native-probability}
evaluates predictive distributions and between-fit dispersion.

\clearpage
\section{Data, visual reduction and common evaluation}
\label{sec:representation}

The four tasks use public DINO-WM PointMaze and Wall trajectories
\citep{zhou2025dinowm}, PhenoCam imagery and greenness products
\citep{ballou2025phenocamimages,zimmerman2025phenocamgreenness,richardson2018phenocam},
and SKIPP'D sky images with photovoltaic power \citep{nie2023skippd}.
PointMaze and Wall use episodes 0--15 for training, 16--19 for validation and
100--149 for testing. PhenoCam uses data from 2018--2020 for training,
2021 for validation and 2023 for testing; SKIPP'D uses data from 2017 for
training, 2018 for validation and 2019 for testing.

\begin{table}[htbp]
\centering\small
\caption{Selected ViBR-WM specifications. $P$ counts candidate predictors
per response. All rows use
regression only, $\kappa=0.01$ and $R^2=0.8$. Coefficients are estimated
for all included terms.}
\label{tab:selected-configurations}
\begin{tabular}{@{}llp{0.38\linewidth}p{0.18\linewidth}@{}}
\toprule
Task & Forecast & Candidate design & Inclusion prior\\
\midrule
PointMaze & Recursive, $m=8$ & $P=107$: two states, actions, state--action products and regional terms & \raggedright All $\pi=1$\tabularnewline[3pt]
Wall & Recursive, $m=6$ & $P=42$: two states and actions & \raggedright All $\pi=1$\tabularnewline[3pt]
PhenoCam & Direct, $m=1$ & $P=14$: 4 visual, 4 Gcc-history and 6 calendar terms & \raggedright Visual $\pi=0.5$; others $\pi=1$\tabularnewline[3pt]
SKIPP'D & Direct, $m=1$ & $P=58$: 30 visual, 16 power-history and 12 calendar terms & \raggedright Visual $\pi=0.5$; others $\pi=1$\tabularnewline
\bottomrule
\end{tabular}
\end{table}

\paragraph{PointMaze and Wall inclusion-prior development.}
\label{sec:movement-prior-context}
An earlier exploratory comparison tested $\pi\in\{0.1,0.25,0.5,1\}$ in
38-predictor PointMaze and 34-predictor Wall models across three
representation seeds. Full inclusion ($\pi=1$) gave the lowest mean
validation normalized root mean squared error (NRMSE) in both tasks
and passed the sampling diagnostics.
Those models forecast visual states and then converted them to physical targets
using separately fitted mappings trained on episodes 0--19;
Wall still used its original position labels.
The final joint-state designs use 107 and 42 predictors, respectively,
and hold $\pi=1$ fixed while comparing representations and regression designs.
The earlier search therefore documents the alternatives explored, rather than
a matched inclusion-prior comparison for the final specifications.

PointMaze and Wall use TS-trained projectors followed by PCA. Both scalar tasks
instead reduce frozen DINOv2 features directly: PhenoCam compresses
three-image patch histories, and SKIPP'D summarizes whole-image and quadrant
features over 16 frames.

\paragraph{PointMaze and Wall visual histories.}
The frozen DINOv2 encoder, a small Vision Transformer with
$14\times14$-pixel patches (ViT-S/14), maps an image to a $196\times384$ patch field
with $14\times14=196$ spatial locations \citep{oquab2024dinov2}.
A learned projector, trained with the TS objective independently for each
repetition, gives a $196\times8$ field
\citep{wang2026straightening}. Channel PCA retains seven components because
the projector's final layer normalization restricts its eight channels to
at most seven independent directions. Within each episode, five interleaved
sequences begin at frames 0, 1, 2, 3 and 4 and advance five frames at a time.
PCA of the centered, flattened five-frame differences determines the projection
directions. Each frame's features are projected onto these directions, centered
using training means and scaled by the sample standard deviations of their five-frame increments.
Channel statistics, projection directions, centers, scales and PCA direction signs use
episodes 0--15.
Both final designs use four coordinates, each summarizing the full spatial field. The task-specific scalar image reductions are given below.

PointMaze and Wall use transition interval $\Delta=5$ frames and two-dimensional actions,
so each action block contains ten values. Within an episode of $T$ raw frames,
20-frame windows start at frames $u=0,\ldots,T-20$. When such a window becomes
training row $i$, its origin is $t_i=u+10$ and its response is
$s_{t_i+5}-s_{t_i}$. The two-state design orders its
predictors as $[s_{t_i},A_{t_i},s_{t_i-5},A_{t_i-10},A_{t_i-5}]$;
the three-state candidate inserts $s_{t_i-10}$ before $s_{t_i-5}$.
These additive designs have $Lm+30$ predictors per response. Forecasts start
at frame 10 with frames 0, 5 and 10 available, proceeding for seven transitions
in Wall and ten in PointMaze.

\paragraph{Reference models and repetition.}
The reference architectures are Temporal Straightening, ConvLSTM, PredRNN2017
and SimVP2022 \citep{wang2026straightening,shi2015convlstm,
wang2017predrnn,gao2022simvp}. Their inputs and fixed physical-target
conversions are specified by task.
Every method has ten complete fits per task, or per site in PhenoCam.
PointMaze and Wall vary the trained visual representation and downstream fits;
the visual representation is shared by ViBR-WM and Temporal Straightening.
PhenoCam and SKIPP'D keep the statistical visual basis fixed and vary
sampling seeds; neural fits vary initialization. Each statistical fit uses four
Markov chain Monte Carlo (MCMC) chains, each a separate sequence of parameter samples.

\paragraph{Sampling qualification.}
MCMC diagnostics assess whether the retained
draws adequately represent the posterior. The statistic $\widehat R$
compares within-chain and between-chain variation; values near one indicate
agreement. Effective sample size (ESS) estimates the number of independent
draws carrying comparable information. Bulk ESS assesses sampling precision
across the main part of the posterior; tail ESS assesses precision near its
lower and upper tails.
The numerical criteria below are those adopted in this study.
PointMaze and Wall parameter checks require finite rank-normalized split
$\widehat R\leq1.05$, bulk and tail ESS at least
400 and verification that fixed indicators retain their specified values.
Splitting each chain into halves helps detect changes within a chain;
rank normalization bases the comparison on sample ranks rather than raw values.
Separate checks of the physical
predictive draws use the same limits and are reported after model selection.
PhenoCam and SKIPP'D require $\widehat R\leq1.01$ and bulk/tail ESS at
least 400 for observation variance and conditional predictive means
\citep{vehtari2021rank}. PhenoCam checks every evaluated origin; SKIPP'D
evaluates conditional predictive means at up to 128 evenly spaced forecast
origins chosen before inspecting the MCMC draws.
Separate scalar checks assess coefficients and the set of included predictors
(support). Binary indicators use split $\widehat R$ on their original zero/one
values, without rank normalization, and bulk ESS,
allowing constant zero or one when every retained draw in every chain agrees;
an identically zero coefficient is allowed for an always-excluded term.
The synthetic diagnostic below uses a stricter indicator-switching rule.
Sampling convergence and predictive calibration are evaluated separately;
interval coverage measures the fraction of observed targets inside the predicted
intervals, and distributional scores assess the predictions against test outcomes.

\paragraph{Scoring and interval interpretation.}
NRMSE divides the root mean squared error (RMSE) for a physical target by
that target's training sample standard deviation. For PointMaze and Wall, scoring first divides each coordinate error by its training
standard deviation, averages squared errors over coordinates and horizons,
and takes the square root to obtain one error score per test trajectory for each fitted model.
These scores are averaged across test trajectories and then across the ten repetitions.
PhenoCam computes NRMSE over test weeks for each site and horizon, then averages
equally over sites, horizons and repetitions. SKIPP'D computes NRMSE over forecast
starting times at each horizon, then averages over horizons and repetitions.
The training standard deviation is coordinate-specific for PointMaze and Wall,
site-specific for PhenoCam and horizon-specific for SKIPP'D.
Bootstrap intervals for differences in mean error use
2,000 resamples and the 2.5th/97.5th percentiles of
ViBR-WM error minus comparator error, without adjustment for multiple comparisons.
Each resample selects data blocks and fitted-model results with replacement;
the same selected data blocks are used across methods.
PointMaze and Wall resample paired ViBR-WM/TS fits together to preserve
their shared visual representations and resample RGB fits independently; scalar tasks resample fits
independently between methods. These intervals condition on fixed environments,
training data, selected specifications and conversions. They exclude uncertainty
from model selection, measurement refitting and new environments.
Task-specific blocks appear below.

\clearpage
\section{Wall: specification and results}
\label{sec:wall-native-final}

\paragraph{Image-aligned position targets.}
The released position labels differ from the positions visible in the source images. Replaying the original
environment with the supplied actions reproduces 815 of 816 training
frames exactly; the remaining mean absolute RGB difference is 0.000160
on a 0--255 scale. Recorded and image-consistent positions differ by up to
33.5941 units.
Each archived episode contains 51 images, giving 816 images in this replay
check. The modeling protocol uses frames 0--49 of each training episode,
giving the 800 position observations used for normalization below.
We retain the original images/actions and derive image-consistent positions
from the environment. RGB predictors are unchanged; position-conditioned
models are refitted using these environment-derived positions.
All methods are scored against the reconstructed positions.

\paragraph{Selected design and inputs.}
Episodes 0--15 are used for training, 16--19 for validation, and the
50 trajectories 100--149 for testing.
Validation selects the two-state, 42-predictor design (NRMSE 0.251163) over
the three-state, 48-predictor design (0.295257).
Four visual and two position coordinates are jointly
predicted, with $2(4+2)+30=42$ regressors per response and 496 overlapping
training windows. Fixed wall geometry adds no regression columns.
All ten fits use four chains of 4,000 iterations, discard 2,000 per chain,
and retain $\pi=1$, $R^2=0.8$ and $\kappa=0.01$. Every fit passes the
parameter, fixed-inclusion and separate predictive-draw checks.

Temporal Straightening receives RGB images, the two observed position coordinates
through its proprioceptive input, and known two-axis actions. Its ten models use
the parameters after 20 training epochs. A separate mapping converts predicted
proprioceptive embeddings, the model's learned vector representations of position,
into positions. This mapping is fitted to embeddings
and positions from the same frames, selected using training episodes 0--11
versus 12--15, refitted on episodes 0--15 and held fixed during evaluation.
The three RGB
architectures use three historical frames to predict three future frames.
They train for 100 epochs; forecasting uses the parameters from the epoch
with the lowest RGB validation loss.
Five-frame-step recursion consumes only their own RGB predictions.
For the RGB models, a separate ridge regression maps DINO features to positions
using image--position pairs from the same frames and the same $64\times64$
image processing. The DINO encoder is held fixed. The coefficient-shrinkage
penalty is $10^{-4}$, selected within episodes 0--15 before refitting on all
of those episodes. On actual
validation images its coordinatewise NRMSE is 0.04418/0.04063.
Clipping and physical conversion are applied only for scoring, never fed
back into RGB recursion. Physical error reflects both the forecasting
network and its fixed measurement.

\paragraph{Common targets and results.}
Panels b--d of Figure~\ref{fig:statistical-method} show the first prescribed Wall test trajectory (episode 100)
and fit (seed 202608271).
From origin frame 10, the four targets are frames 15/25/35/45, or
5/15/25/35 raw frames ahead. ViBR-WM uses history frames 5/10;
TS and RGB predictors receive frames 0/5/10. Physical normalization
uses all 800 corrected position frames in the 16 training trajectories,
with sample standard deviations $[18.8223456188,18.2250273512]$.
Each trajectory's overall NRMSE is the root mean square (RMS) of standardized
prediction errors over four targets and two coordinates; scores are then
averaged equally over the 50 trajectories and ten fits.
Figure~\ref{fig:statistical-method}a gives the overall means and sample standard deviations (SDs);
Tables~\ref{tab:wall-native-horizons} and~\ref{tab:wall-native-repeats}
report every horizon and fit. Comparing fits with the same repetition index,
ViBR-WM has lower overall NRMSE than each comparator in all ten comparisons.
These comparisons are descriptive; only ViBR-WM and TS share a trained
representation within each index. Against TS at the farthest target,
ViBR-WM has lower error in nine comparisons and higher error in one;
it has lower error in all ten comparisons at every other target.

\paragraph{Whole-trajectory and fit uncertainty.}
The bootstrap jointly resamples whole trajectories and fitted models in
2,000 draws.
Whole-trajectory indices are shared across all methods, fits and horizons.
Each sampled ViBR-WM fit is paired with the TS fit using the same trained visual representation;
ConvLSTM, PredRNN and SimVP fit indices are sampled independently.
The four overall ViBR-WM-minus-comparator 2.5th/97.5th percentile
intervals are negative. TS target-frame 35 and 45 intervals include zero,
while intervals for all four horizons against each of the three RGB methods lie
below zero (Table~\ref{tab:wall-native-intervals}). A centering shift is the bootstrap-mean
error difference minus the original observed error difference;
the intervals use the original bootstrap draws without subtracting this shift.
The intervals condition on the reconstructed positions and use the common
conditions in Section~\ref{sec:representation}, with trajectories treated
as independent resampling units.
ViBR-WM's posterior predictive calibration is reported below.
\input{generated/wall_native_final_supplement.tex}
\FloatBarrier

\clearpage
\section{PointMaze: specification and results}
\label{sec:pointmaze-native-final}

For PointMaze, write $s_t=(z_t^\top,p_t^\top,v_t^\top)^\top\in\mathbb R^8$,
where the local symbols $z_t$, $p_t$ and $v_t$ denote its four standardized
visual, two position and two velocity coordinates. Physical coordinates are
centered and scaled using all frames of training episodes 0--15; scales are
sample standard deviations. Let
$\bar a_t\in\mathbb R^2$ be the mean of the supplied five-step action block.
The additive design retains two eight-dimensional states separated by five frames and thirty
ordered action values, giving 46 predictors per response. A first extension
adds all $s_{tj}\bar a_{tk}$ products, $j=1,\ldots,8$, $k=1,2$, giving 62.

A second extension lets motion relations vary smoothly with position.
Let $Q_{\alpha,k}$ be quantile $\alpha$ of position coordinate $k$ in the
training transition rows. Pairing each coordinate's 25th, 50th and 75th
percentiles gives nine region centers $c_r\in\mathbb R^2$.
Let $d=(d_1,d_2)$, where $d_k=(Q_{.75,k}-Q_{.25,k})/2$ controls how quickly
each regional weight decreases along coordinate $k$. Define
\begin{equation}
 g_r(p)=\operatorname{exp}\!\left\{-\frac12\sum_{k=1}^2
             \left(\frac{p_k-c_{rk}}{d_k}\right)^2\right\},
 \qquad r=1,\ldots,9.
\end{equation}
Each weight $g_r(p)$ is largest at its region center and decreases with
distance; weights are not normalized to sum to one. Append
$g_r(p_t)(1,v_{tx},v_{ty},\bar a_{tx},\bar a_{ty})$ for each region:
$46+16+9\times5=107$ predictors. These products give each region an intercept
and its own velocity and action effects. New columns are standardized using training
rows only; the existing additive columns and response normalization are
retained. Joint regression predicts the eight-dimensional five-frame
increment $\Delta s_t=s_{t+5}-s_t$ with a full disturbance covariance;
recursive prediction adds each sampled increment to its current state.

All eleven candidates use training episodes 0--15, validation episodes
16--19 and four chains of 4,000 iterations with 2,000 discarded.
Table~\ref{tab:pointmaze-selection} lists the validation candidates.
The validation score takes the RMS of standardized prediction errors across
horizons and physical coordinates within each validation trajectory,
then averages the four trajectories. All candidates pass their
parameter/prediction checks. The minimum selects $q=4$, two histories,
107 predictors, $\kappa=0.01$, $R^2=0.8$, fixed inclusion and the
regional scales $d$ defined above. Nine further fits complete ten repetitions of this specification.
Each learns its own visual reduction, scaling and regional centers.
All eleven candidates fix inclusion at one, so validation compares state
representations and regression designs.

\begin{table}[htbp]
\centering\small
\caption{PointMaze validation candidates. $q$ is visual dimension, $L$
history length and $P$ predictors per response.}
\label{tab:pointmaze-selection}
\begin{tabular}{lrrrrr}
\toprule
Design & $q$ & $L$ & $P$ & $\kappa$ & NRMSE\\
\midrule
Additive & 4 & 2 & 46 & 0.01 & 0.640590\\
Additive & 8 & 2 & 54 & 0.01 & 0.658151\\
Additive & 16 & 2 & 70 & 0.01 & 0.670158\\
Additive & 4 & 3 & 54 & 0.01 & 0.659372\\
Additive & 8 & 3 & 66 & 0.01 & 0.674185\\
Additive & 16 & 3 & 90 & 0.01 & 0.691281\\
State--action products & 4 & 2 & 62 & 0.01 & 0.608455\\
Regional (selected) & 4 & 2 & 107 & 0.01 & 0.572091\\
Regional & 4 & 2 & 107 & 129.6 & 0.574630\\
Regional & 4 & 2 & 107 & 1296 & 0.596578\\
Regional, scales $d/2$ & 4 & 2 & 107 & 0.01 & 0.573473\\
\bottomrule
\end{tabular}
\end{table}

During test evaluation, each fit's visual reduction, scaling and regional centers
remain fixed. Every recursive
sample recomputes state--action products and regional weights $g_r(p)$ from its own predicted state and the
supplied actions.

\subsection*{Forecast evaluation}
All ten ViBR-WM fits pass the parameter and prediction checks.
Across ten repetitions of the selected specification, mean validation NRMSE
is 0.5911755 with between-fit standard deviation 0.0164355.
From origin frame 10, forecasts are evaluated at frames 15, 25, 35 and 60,
corresponding to 5, 15, 25 and 50 raw frames ahead.
The selected models predict episodes 100--149. Every method is scored on
the same 50 trajectories, four horizons and training coordinate scales.
For each trajectory, take the root mean squared normalized error across
the four horizons and four physical coordinates, then average equally
over trajectories and the ten fits.

All three RGB families have ten fits. Within each training window,
three RGB frames at relative indices 0, 5 and 10 predict three future frames.
Each fit trains for 100 epochs on episodes 0--15; forecasting uses the parameters
from the epoch with the lowest RGB validation loss on episodes 16--19.
Five-frame-step recursion uses each network's
own predictions.
Temporal Straightening uses a predictor trained for 20 epochs on images,
observed position and velocity, and actions. Its predicted proprioceptive
embeddings are converted into position and velocity by a separate mapping
fitted to embeddings and physical states from the same time points.

The RGB scoring conversion uses frozen DINO features from frame pairs
$(t-5,t)$ at the same $64\times64$ resolution as network predictions.
Its ridge penalty is selected within training episodes 0--11 versus
12--15, then the measurement is refitted on all 0--15. The chosen
penalty is 100. Targets are
the recorded instantaneous $(x,y,v_x,v_y)$, not a finite-difference average
velocity. The first pair is observed frame 10 and predicted frame 15;
later pairs contain predictions only. RGB clipping is for measurement
only, not recursion. On actual validation images, coordinatewise NRMSE
is 0.0886/0.0636/0.8652/0.5896 for $x,y,v_x,v_y$, respectively.
This fixed measurement error remains part of the system comparison.

Comparing fits with the same repetition index, ViBR-WM's overall error is
lower than each comparator in all ten comparisons. At every target frame it is lower
in all ten comparisons against the RGB methods; against Temporal
Straightening, its frame-35 error is lower in nine comparisons and higher in one. Tables~\ref{tab:pointmaze-native-horizons}
and~\ref{tab:pointmaze-native-repeats} retain every horizon and fit.

\paragraph{Whole-trajectory and fit resampling.}
The bootstrap uses 2,000 draws. Each draw samples 50
whole trajectories with replacement, shared across all methods, fits and
horizons. ViBR-WM and Temporal Straightening share ten sampled fit indices
because each repetition uses a common epoch-20 representation. Each RGB family
instead samples its ten fit indices independently. The same draws produce the
overall and four horizon contrasts, retaining the trajectory RMS rule.

All four overall ViBR-WM-minus-comparator 2.5th/97.5th percentile
intervals lie below zero (Table~\ref{tab:pointmaze-native-intervals}).
All RGB horizon intervals also lie below zero. Temporal Straightening intervals
lie below zero at frames 15, 25 and 60; its frame-35 interval is $[-0.1208,0.0122]$.
The intervals use the conditions in Section~\ref{sec:representation}
and assume independent trajectory clusters. Their nominal repeated-sampling
coverage has not been established.
\input{generated/pointmaze_native_final_supplement.tex}
\FloatBarrier

\clearpage
\section{PhenoCam and SKIPP'D: specifications and results}
\label{sec:scalar-native-final}

Both tasks predict scalar physical targets from visual features, numerical
histories and known calendar covariates. Comparator inputs are detailed
below; numerical/calendar augmentations are evaluated in
Section~\ref{sec:input-sensitivity}.
Each site and horizon uses a separate scalar response regression.
Besides uniform inclusion priors, the regression-only candidates also test
a mixed prior that keeps numerical and calendar terms included while
assigning each visual term prior inclusion probability $0.5$; all included
coefficients remain estimated.
PhenoCam uses data from 2018--2020 for training and 2021 for validation;
SKIPP'D uses data from 2017 for training and 2018 for validation.
Both tasks compare ten fits of all five methods. Test evaluation keeps
the selected basis, scaling and prior fixed.

\subsection{PhenoCam}
\label{sec:phenocam-specification}

\paragraph{Regression design and model selection.}
In PhenoCam the response $g_t$ is green chromatic coordinate (Gcc) at week
$t$, green intensity divided by total red, green and blue intensity.
Let $\tau$ be the target week and $t=\tau-h$ the forecast origin, with
$h\in\{1,4,13,26\}$ weeks. The four
historical Gcc regressors are
\begin{equation}
 (g_{t-2},g_{t-1},g_t,g_{\tau-52}).
\end{equation}
The last regressor aligns to the target's week in the preceding year;
all four observations are available at the forecast origin. Scaling uses
the training target mean and sample standard deviation. The weekly clock
retains missing weeks: each input uses the most recent observation no
later than its nominal week. For candidates with a latent state, missing
observations retain their calendar positions during filtering.

At each spatial patch, channel PCA reduces 384 frozen DINOv2 features
to eight, without a TS projector. The image fields at weeks $t-2,t-1,t$,
each of size $8\times14\times14$, are flattened and concatenated into one
history vector. A second PCA reduces this vector; its first $K$ components give $K$
image regressors. Six
further regressors are sine/cosine pairs with periods of 365.2425 days,
half that period and one third of that period, evaluated at the Sunday
ending the target week.
Thus $K\in\{1,2,4,8,16,32,64,96\}$ gives $K+10$ predictors;
the mixed prior fixes the ten numerical/calendar inclusion indicators.
The stateful candidates additionally filter a random local level, with or
without a random 52-week seasonal component, using the observed Gcc history. The
regression-only candidate has no such state update: numerical history enters
only through its four regressors.

The PhenoCam validation search compares 33 declared candidates:
30 pass the predictive-sampling checks at every combination of site and horizon, and three do
not. The selected regression-only mixed-prior model has $K=4$ and
$\kappa=0.01$, $R^2=0.8$, with mean NRMSE 0.3581110 in the initial selection
fit across two sites and four horizons. All eight selected predictive, support and coefficient
diagnostic groups pass. The same mixed prior at
$K=1,2,4,8,16,32,64,96$ gives means
0.3641, 0.3642, 0.3581, 0.3585, 0.3607, 0.3618, 0.3636 and 0.3636;
increasing $K$ beyond four does not reduce validation error. Retaining every term at
$K=96$ gives 0.5108 and passes qualification. After nine additional fits,
averaging validation scores over all ten repetitions of the selected specification
gives NRMSE 0.3580941, with all 80 site--horizon--fit
predictive, support and coefficient diagnostic groups passing. Each PhenoCam fit uses four chains of 48,000 iterations, discarding
16,000 per chain.

\paragraph{Structural and selection sensitivity.}
Table~\ref{tab:phenocam-design-sensitivity} extracts controlled contrasts
from the same 33-candidate validation search. At $K=16$, keeping the
regression design and priors fixed while removing stochastic seasonal
and level states improves validation NRMSE. Deterministic calendar terms
remain in every row. No cycle-state comparison is included in this search.
At fixed $K=4,16,96$, selecting visual terms improves validation error
relative to retaining all terms. The search does not include matched fits
without visual regressors.

\begin{table}[htbp]
\centering\small
\caption{PhenoCam validation sensitivity within the declared search.
All candidates use the same 2018--2020 training data, 2021 validation
targets, $\kappa=0.01$ and $R^2=0.8$. NRMSE averages two sites and four
horizons. ``Pass'' counts the site--horizon combinations passing
predictive-sampling checks out of eight;
no aggregate is reported for an unqualified candidate.}
\label{tab:phenocam-design-sensitivity}
\begin{tabular}{@{}lrllr@{}}
\toprule
Structural states & $K$ & Inclusion & Pass & NRMSE\\
\midrule
Level + seasonal & 16 & All fixed & 8/8 & 0.506103\\
Level & 16 & All fixed & 8/8 & 0.456370\\
None & 16 & All fixed & 8/8 & 0.385197\\
None & 16 & Visual selection & 8/8 & 0.360727\\
None & 16 & All selectable & 7/8 & Unqualified\\
\midrule
None & 4 & All fixed & 8/8 & 0.363566\\
None (selected) & 4 & Visual selection & 8/8 & 0.358111\\
None & 96 & All fixed & 8/8 & 0.510763\\
None & 96 & Visual selection & 8/8 & 0.363633\\
\bottomrule
\end{tabular}

\smallskip
\parbox{\linewidth}{\small $K$ is the number of visual regressors;
the ten numerical/calendar regressors are unchanged. ``All fixed'' sets
every $\pi=1$; ``Visual selection'' sets visual $\pi=0.5$ and other
$\pi=1$; ``All selectable'' sets every $\pi=0.5$. Each site--horizon
candidate is one four-chain fit.}
\end{table}

\paragraph{Comparator inputs, objectives and conversions.}
The 60 RGB fits comprise ten initializations for each of three methods at
each site. They train for 100 epochs, batch size four, without early stopping,
on 155 three-future-frame blocks per site with targets in 2018--2020.
Forecasts use the network parameters from the epoch with the lowest
three-frame RGB prediction loss on the 2021 validation data.
ConvLSTM uses a paper-based independent
encoder--forecaster with peephole cells, binary cross-entropy and RMSProp
\citep{shi2015convlstm}; PredRNN2017 uses the OpenSTL implementation,
$L_1+L_2$ loss, Adam and ordinary scheduled sampling with
reversed-sequence augmentation
\citep{wang2017predrnn,tan2023openstl}. SimVP2022 uses
mean squared error, Adam and a one-cycle schedule
\citep{gao2022simvp}. Learning rates are 0.001, 0.001 and 0.01,
respectively. The RGB channel/size interface is adapted to the dataset.

Each of three historical images is mapped to a $64\times64$ grid over
the bounding rectangle of the vegetation region of interest (ROI).
Each occupied cell stores the mean
RGB values of its original ROI pixels; empty cells are filled with gray.
ConvLSTM and PredRNN continue their forecast state to 26 weeks;
SimVP recursively uses its own three-frame blocks.
To convert a predicted grid to Gcc, each cell's RGB values are weighted by
its fixed count of original ROI pixels. The weighted green sum is divided
by the weighted sum across all three channels; empty cells have zero weight.
RGB values are clipped to the physical
range only for this measurement, not for network feedback.

The 20 Temporal Straightening fits train on 2018--2020 for 20 epochs
and select the minimum 2021 total validation loss, retaining
the earlier epoch in a tie \citep{wang2026straightening}. They use
ROI-cropped $224\times224$ images represented by 196 spatial patch vectors.
Three historical RGB images enter the visual input, and the corresponding
Gcc observations enter the proprioceptive input; future actions are zero.

Let $f(g)\in\mathbb R^{10}$ be the learned encoder that maps a Gcc value to
its proprioceptive embedding, including training normalization, projection
and layer normalization. This encoder is held fixed during physical conversion.
For a predicted proprioceptive embedding $z$, the physical conversion is
\begin{equation}
 \widehat g=\underset{g\in[0,1]}{\operatorname{argmin}}\,
                 \|f(g)-z\|_2^2.
\end{equation}
We evaluate all stationary points of this one-dimensional squared-distance
objective and its endpoints $g=0,1$, then choose the minimum with deterministic
tie-breaking.
The rule was fixed before the present 2023 evaluation. All 20 encoders pass the
157-observation round-trip check: encode each training Gcc value, invert the
embedding and compare the recovered Gcc with the original. The largest normalized
error is $2.78\times10^{-7}$. Conversion is applied after generating the feature forecasts.
A conditional mean in latent space need not lie on the encoded curve;
closest-point conversion therefore need not minimize physical MSE.
The earlier training-only ridge conversion gives overall NRMSE 1.0261896
and is reported as a conversion-sensitivity result.

\paragraph{Test data and results.}
All five methods score the same 48, 45, 36 and 24 weekly targets per site
at horizons 1, 4, 13 and 26 in 2023. Dates use ISO week numbering
(Monday-to-Sunday weeks), with missing week 14 retained on the calendar.
Available 2022 images provide historical context.
Images are aligned by their actual ISO year/week to the forecast starting times,
histories, targets and truths. Original full-training Gcc standard deviations are
0.0307122088532 and 0.0353830791780 at Mead1 and Mead3, from 157
observations each. Within a site/horizon/fit, NRMSE is weekly RMSE divided
by this scale; ten fits, four horizons and two fixed sites are averaged
equally. Each fit is scored before averaging.

Overall means are ViBR-WM 0.4367004, Temporal Straightening 1.6929093,
ConvLSTM 0.9718132, PredRNN 1.0281311 and SimVP 0.8798639.
Their across-repeat sample standard deviations are 0.0000433, 1.1282623,
0.0253018, 0.0927177 and 0.1170443. Comparing fits with the same repetition
index, ViBR-WM has lower overall error than each comparator in all ten comparisons.
Relative to Temporal Straightening, ConvLSTM, PredRNN and SimVP, respectively,
its mean error is lower by 74.20\%, 55.06\%, 57.52\% and 50.37\%.
At one week ahead, ConvLSTM and PredRNN
are better on average: 0.2997 and 0.3029 versus ViBR-WM 0.3213.
ViBR-WM has lower error in one and two of those ten one-week comparisons,
respectively. The same ViBR-WM fits
are used in the sensitivity comparison in Section~\ref{sec:input-sensitivity}.
For ViBR-WM's posterior predictive distribution, the 95\% intervals contain
78.8715\% of observed targets. The normalized continuous ranked probability
score (CRPS) is 0.2374053 and mean normalized 95\% interval width is 1.0958442.
Section~\ref{sec:native-probability} defines these scores and compares them
with the variation in predicted means across ten fits.

\paragraph{Temporal-block uncertainty.}
The analysis protocol was fixed after the point-score comparison and before
bootstrap execution. It uses 13-week blocks in the primary analysis and
26-week blocks in a sensitivity analysis, each with 2,000 bootstrap draws
and seed 202609405. Consecutive blocks, without wrapping from year end to
year start, are sampled on the original 52-week calendar, concatenated and
truncated to 52 positions. All methods, fits, sites and horizons share the
same calendar weights, meaning that a given week occurs equally often for
every method, fit, site and horizon. Missing observations remain masked rather than
compressing the calendar. A draw with no evaluable targets for a site/horizon combination
is redrawn; 21 such draws occur for 13 weeks and none for 26 weeks.
Within a method, ten repeat indices are drawn with replacement jointly
across both sites and all horizons; indices are independent between methods.
The analysis uses the saved forecasts and conversions.
Both analyses give negative overall ViBR-WM-minus-comparator intervals
for all four comparators (Table~\ref{tab:phenocam-native-block-intervals}).
At 4, 13 and 26 weeks ahead, all four comparator intervals lie below zero
under both block lengths. At one week ahead, all four intervals from the
13-week-block analysis include zero. With 26-week blocks, the one-week-ahead
interval lies below zero against TS, above zero against PredRNN, and includes
zero against ConvLSTM and SimVP.

The intervals are unadjusted for multiple comparisons and condition on two
fixed sites, one year, the selected models and frozen conversion. ViBR-WM repetitions vary MCMC
randomness on one fixed DINO/PCA/design; neural repetitions vary training
initialization. With only 24 targets at horizon 26, the observations span
$24/13\approx1.85$ primary or $24/26\approx0.92$ sensitivity block lengths. Noncircular blocks also
underweight endpoint weeks; the percentile intervals use the bootstrap draws
without correcting any shift between the bootstrap mean and the observed
error difference. Agreement between the two block lengths does not
establish nominal coverage or capture all seasonal dependence.
\input{generated/phenocam_native_final_supplement.tex}
\FloatBarrier

\subsection{SKIPP'D}
\label{sec:sky-native-returned}

\paragraph{Regression design and model selection.}
For SKIPP'D, 16 one-minute sky images and corresponding power observations
end at origin $t$; horizons are 5, 15, 30 and 60 minutes. Visual groups comprise
the DINO class token, which summarizes the whole image, and features pooled
within each of four spatial quadrants. Each of these five groups retains
$K$ principal components. Each component contributes its value at the origin,
its 16-frame mean and the slope of a least-squares line through its 16 values.
Thus five groups, $K$ components and three summaries give $15K$ visual regressors.
The 16 observed power values and 12 target-calendar values complete the
design. Calendar terms comprise sine/cosine pairs for one, two and three
cycles per day and per year, evaluated at the target time in local Los Angeles time.
Consequently $K=2,4,8$ yields
58, 88 and 148 predictors; the mixed prior fixes the 28 numerical/calendar
inclusion indicators. These horizon-specific regressions use deterministic
calendar regressors and constant coefficients; the selected specifications
have no stochastic structural state.

The selected SKIPP'D design uses $K=2$, giving 58 predictors, with
$\kappa=0.01$, $R^2=0.8$ and the mixed inclusion prior above. Its ten
four-chain fits retain 20,000 of 24,000 iterations per chain. The mean
2018 validation score is 0.4270779. All 40 fit--horizon predictive diagnostic
groups pass. Coefficient/support interpretation passes in 39 groups;
fit 5 at 15 minutes is used for prediction but not coefficient interpretation:
one visual coefficient's tail ESS is undefined, failing the requirement for
a finite diagnostic value.

\paragraph{Inputs, conversions and results.}
All methods use the same 3,124 origins, physical targets and 2017
target scales at 5, 15, 30 and 60 minutes. Temporal Straightening receives
16 RGB/PV observations through its visual/proprioceptive interfaces and zero actions;
the RGB predictors receive the 16 images. For TS, a fixed regression maps
predicted proprioceptive embeddings to power. For the RGB models, another
fixed regression maps DINO features of predicted images to power. These
regressions are fitted to 2017 embeddings or image features paired with
power measurements at the same times.
ViBR-WM also uses historical power and the specified
calendar terms.

All five methods have ten complete fits. For each fit and horizon, NRMSE is
the RMSE across forecast starting times divided by the corresponding 2017
training-target sample standard deviation. Scores are then averaged equally
over horizons and fits.
Table~\ref{tab:skippd-native-final-2019} gives overall means and sample SDs;
Tables~\ref{tab:skippd-native-horizons} and~\ref{tab:skippd-native-repeats}
report all horizons and repetitions. ViBR-WM has the lowest overall mean
NRMSE (0.2180), compared with 0.4290 for PredRNN, the lowest-error comparator.
ViBR-WM's mean error is lower than every comparator at each evaluated horizon.
ViBR-WM repeats vary Monte Carlo seeds on a fixed basis and design.
ViBR-WM's posterior predictive 95\% intervals contain 97.43\% of observed
targets, above their nominal 95\% level. Intervals based only on the spread
of ten fitted mean predictions contain 1.99\%
(Section~\ref{sec:native-probability}).

\paragraph{Calendar-week and fit uncertainty.}
The 2,000-draw bootstrap resamples the 27 whole calendar-week blocks
jointly across methods and horizons, and resamples ten complete fits
independently within each method. All four overall ViBR-WM-minus-comparator
95\% percentile intervals are negative
(Table~\ref{tab:skippd-native-contrasts}). Only overall intervals were
computed; horizon means are descriptive. These unadjusted intervals
condition on the fixed training data,
selected specifications, conversions and the 2019 test set;
they do not include model-selection or between-year uncertainty.

\begingroup
\setlength{\floatsep}{6pt}
\setlength{\textfloatsep}{8pt}
\setlength{\intextsep}{6pt}
\input{generated/skippd_native_final_main.tex}
\input{generated/skippd_native_final_supplement.tex}
\FloatBarrier
\endgroup

\clearpage
\section{Predictive distributions and repeat-fit dispersion}
\label{sec:native-probability}

We evaluate two distinct distributions: each ViBR-WM fit's posterior
predictive distribution, and the empirical distribution of ten predicted
means from repeated fits of each method. The first quantifies uncertainty within a fitted
model; the second measures variation between fitted-model predictions.

For a standardized scalar target $y$ and an empirical distribution with
$M$ equally weighted values $x_1,\ldots,x_M$, the continuous ranked
probability score (CRPS)
\citep{gneiting2007proper} is
\[
 \operatorname{CRPS}
 =\frac{1}{M}\sum_i|x_i-y|
 -\frac{1}{2M^2}\sum_{i,j}|x_i-x_j|.
\]
The energy score replaces absolute distances by Euclidean distances for
the joint position or position--velocity vector. Lower values indicate
better distributional predictions. These are exact scores of the finite empirical distribution,
including self-pairs in the double sum. Coordinates are scaled by their fixed
training standard deviations. For PointMaze and Wall, marginal scores and coverage are
averaged equally over coordinates, trajectories and horizons. PhenoCam weeks are averaged
within each site/horizon, then the four horizons and two sites equally;
unequal week counts do not reweight sites or horizons. SKIPP'D averages its
3,124 common origins within each horizon, then four horizons equally.
For scalar targets, the energy score equals CRPS and is omitted.
For PointMaze and Wall, the energy score assesses coordinates jointly at each horizon.
Central intervals use linearly interpolated empirical quantiles.

\paragraph{Within-fit posterior prediction.}
Table~\ref{tab:native-statistical-posterior-probability} scores
ViBR-WM's posterior predictive distribution within each fit,
then averages those scores over fits and the same evaluation units. In the
original PointMaze and Wall evaluation, marginal CRPS and central intervals were
computed from all 8,000 pooled posterior predictive draws per fit; energy
used 256 draws, 64 equally spaced within each of four chains.
Scores are aggregated from saved per-fit metrics.
Wall interval endpoints are unavailable, so its quantiles
are not independently recomputed. Posterior predictive 95\% coverage is 78.87\% in PhenoCam,
93.66\% in PointMaze, 89.03\% in Wall and 97.43\% in SKIPP'D.
These intervals are computed from posterior predictive draws without
subsequent recalibration of their width or coverage.

\paragraph{Between-fit dispersion.}
The ten saved mean predictions at each target define one equally weighted
empirical distribution per method ($M=10$). No residual noise or
calibration scale is added. ViBR-WM's ten-mean intervals cover only 0.28\%
and 1.99\% of PhenoCam and SKIPP'D targets, respectively, compared with
posterior predictive coverage of 78.87\% and 97.43\%.
In PhenoCam and SKIPP'D, the fixed visual
features mean that variation between ViBR-WM predictions largely reflects
finite posterior sampling; neural fits vary with initialization.
PointMaze and Wall fits also vary with their independently trained upstream
representations. ViBR-WM has the lowest empirical CRPS in PhenoCam and SKIPP'D.
On PointMaze and Wall, ViBR-WM has lower CRPS and energy scores than ConvLSTM,
PredRNN and SimVP, while Temporal Straightening has the lowest scores. Empirical
95\% coverage is below the nominal 95\% level for all five methods on all four tasks
(Table~\ref{tab:native-point-ensemble-probability}).

\input{generated/native_point_ensemble_probability.tex}
\FloatBarrier

\clearpage
\section{Input and conversion sensitivity}
\label{sec:input-sensitivity}

Sensitivity comparisons on the PointMaze and PhenoCam test sets
evaluate task-specific changes to numerical-history inputs, training objectives
based on physical targets, or mappings from model outputs to physical targets.
These changes differ across methods and tasks.
Table~\ref{tab:augmented-sensitivity} reports the resulting PointMaze and
PhenoCam scores with unchanged ViBR-WM fits.

For PointMaze, the three video-model adaptations replace RGB inputs with
three eight-channel feature fields produced by DINOv2 and the TS-trained
projector, together with actions and the four position--velocity coordinates.
They jointly predict standardized visual and physical channels using mean
squared error, with the training epoch selected by validation NRMSE on the
physical coordinates.
The TS sensitivity row keeps its forecasts but selects the physical-conversion
ridge penalty on episodes 16--19 after fitting on 0--15; the main comparison
selects this penalty within 0--15.
For PhenoCam, all four adaptations add a recurrent encoder of Gcc history,
missing-observation flags and dates. A horizon-specific Gcc prediction head
combines the recurrent encoder's output with visual features and six target-calendar terms.
The video models minimize standardized Gcc mean squared error;
the TS adaptation adds this error to its latent-prediction and curvature losses.

\begin{table}[htbp]
\centering\small
\caption{Sensitivity analysis: NRMSE under task-specific input, training or
readout adaptations.}
\label{tab:augmented-sensitivity}
\begin{tabular}{lrr}
\toprule
Method & PointMaze & PhenoCam\\
\midrule
ViBR-WM (unchanged) & 0.560177 & 0.436700\\
Temporal Straightening adaptation & 0.669387 & 0.327527\\
ConvLSTM adaptation & 0.543618 & 0.533886\\
PredRNN adaptation & 0.505283 & 0.354508\\
SimVP adaptation & 0.529514 & 0.342000\\
\bottomrule
\end{tabular}
\end{table}

The adapted ConvLSTM, PredRNN and SimVP models have lower mean NRMSE than ViBR-WM
on PointMaze, whereas the TS adaptation has higher mean NRMSE. In PhenoCam,
the TS, PredRNN and SimVP adaptations outperform ViBR-WM; the respective
ViBR-WM-minus-comparator 95\% bootstrap intervals are $[0.0670,0.1576]$,
$[0.0312,0.1353]$ and $[0.0469,0.1453]$. The ConvLSTM interval
$[-0.2345,0.0351]$ contains zero. These intervals use 2,000 draws of
13-week noncircular calendar blocks, with shared calendar weights and fits
resampled independently by method and site, jointly across horizons.

To isolate conversion effects, an earlier PhenoCam TS ridge conversion
is compared with the nearest-point rule on the same latent forecasts.
The ridge conversion has NRMSE 1.026190, versus
1.692909 for the nearest-point conversion using the fitted TS encoder.
This readout comparison holds the TS forecasts fixed. The video-model
adaptations and the PhenoCam TS adaptation in
Table~\ref{tab:augmented-sensitivity} change several input or training
choices together. Their results cannot isolate the contribution of
any one additional input, prediction head or optimization objective.
Likewise, PointMaze's conversion error on actual images is a diagnostic
of the measurement procedure, not a proven lower bound on forecast error.

\clearpage
\section{Controlled diagnostics}
\label{sec:synthetic}

\subsection{Known-support dynamics}
\label{sec:known-support-dynamics}
This diagnostic tests whether variable selection identifies inputs whose
true coefficients are nonzero, separately from forecast accuracy.
There are four state coordinates, two actions and two additional predictors
with zero true coefficients, giving eight inputs per response. The dynamics are
$s_{t+1}=Fs_t+Ga_t+\eta_t$, with
$F=\operatorname{diag}(0.72,0.76,0.70,0.78)$,
$G=[(0.30,0.18);(-0.20,0.28);(0.25,-0.18);(-0.22,-0.26)]$
and $Q_{ij}=0.10^2\,0.5^{|i-j|}$ for $\eta_t\sim\mathcal N(0,Q)$.
The increment coefficient matrix $[F-I\mid G\mid0]$ has 12 nonzero and
20 zero entries; $I$ is the $4\times4$ identity and the final block is
$4\times2$ zeros. Actions follow first-order autoregression with coefficient
0.4, so their current values depend on their previous values plus new noise.
The two independent zero-effect predictors each follow first-order autoregression
with coefficient 0.8.
Four data seeds (2026100101--2026100104) each supply
six training episodes of 40 transitions, two validation episodes of five
transitions and 20 test episodes of five transitions, after discarding the
first 100 simulated transitions to reduce the influence of initialization.
We compare joint-response regressions that allow correlated disturbances
across responses with independent-response regressions that treat those
disturbances as independent, each with and without variable selection,
at horizons 1, 3 and 5.

The prespecified variable-inclusion diagnostic requires each selectable
indicator to visit both zero and one within every chain, a difference of at most
0.10 between the largest and smallest chainwise inclusion probabilities,
and the applicable continuous-parameter checks.
Applying a 0.5 threshold to pooled posterior inclusion probabilities correctly
classifies every coefficient as zero or nonzero for
4/4 seeds in both joint- and independent-response models, but strict inclusion-probability
diagnostics pass in 0/4 for both. Thus these matches do not establish
reliable estimation of which predictors have nonzero coefficients.
A sensitivity diagnostic that permits an indicator to remain at zero in all
chains or at one in all chains passes 4/4 in each family, but was defined
after this failure and is post hoc. Relative to retaining every predictor,
variable selection reduces the mean
energy score by 0.004/0.011/0.013 at horizons 1/3/5 for the joint model
and 0.003/0.010/0.013 for independent responses. With selection enabled in both
models, joint-response modeling reduces the mean energy score by 0.001/0.002/0.002 relative
to independent responses at horizons 1/3/5.

\subsection{Fixed-history, multiple-future prediction}
\label{sec:multiple-future-prediction}
To assess uncertainty over alternative futures, this diagnostic generates
repeated futures conditional on the same observed history and action schedule.
Each future comes from one of two equally likely branches, representing
different future-trajectory distributions whose identity is unobserved at
the forecast origin.
We repeat it over 20 data seeds (2026100101--2026100120). Let $\delta$ be the
difference between the two branch-specific five-step mean trajectories,
each a 20-dimensional vector of four responses at five steps, and let $V_5$
be their common within-branch covariance. The separation is
$D_5=(\delta^\top V_5^{-1}\delta)^{1/2}$.
Table~\ref{tab:multi-future-compact} reports the whole-path energy score after transforming
each 20-coordinate path by $L^{-1}$, where $V_5=LL^\top$ is its Cholesky
factorization. This whitening places within-branch variation on a common
unit-covariance scale; horizon scores and seeds accompany the saved summaries.
This diagnostic uses joint-response regression with every term included.
The mean-only distribution places all mass on the regression model's posterior predictive
mean trajectory. The empirical-path distribution subtracts the average training
trajectory from each training path and adds that same predictive mean,
preserving dependence within paths. The moment-matched Gaussian uses the
generating distribution's true mean and covariance; the two-branch mixture
uses its actual branches. These are oracle comparisons, meaning that they
use the known data-generating distribution rather than estimating it from training samples.

\begin{table}[htbp]
\centering\small
\caption{Multiple-future diagnostic: mean whole-path energy score for five-step
trajectories, $\pm$ the standard deviation across 20 data seeds; lower is better.
$D_5$ measures separation between the two branch means relative to
within-branch variation.
The regression component is evaluated without a visual encoder.}
\label{tab:multi-future-compact}
\begin{tabular}{lccc}
\toprule
Predictive distribution & $D_5=0$ & $D_5=2$ & $D_5=4$\\
\midrule
Regression posterior & $3.160\pm0.017$ & $3.235\pm0.016$ & $3.445\pm0.017$\\
Regression mean only & $4.465\pm0.024$ & $4.572\pm0.023$ & $4.882\pm0.024$\\
Recentered empirical paths & $3.182\pm0.017$ & $3.257\pm0.016$ & $3.468\pm0.017$\\
Moment-matched Gaussian & $3.135\pm0.013$ & $3.210\pm0.013$ & $3.421\pm0.013$\\
True two-branch mixture & $3.135\pm0.013$ & $3.210\pm0.013$ & $3.421\pm0.013$\\
\bottomrule
\end{tabular}
\end{table}

The regression posterior has a lower mean energy score than mean-only and recentered-path
distributions at all three separations. Empirical-path minus posterior energy-score differences
are $+0.022013$, $+0.022471$ and $+0.022214$ for $D_5=0,2,4$, positive in all 20 seeds.
True-mixture minus moment-matched-Gaussian energy-score differences are 0, $-0.000113$ and
$-0.000709$; at $D_5=4$, 10,000 data-seed bootstrap draws give the 95\% interval
$[-0.000867,-0.000565]$. The table's three-decimal rounding hides these small
differences. The oracle moment-matched Gaussian nearly matches the true-mixture
energy scores without recovering the branches.
\FloatBarrier

\suppressfloats[t]
\clearpage
\section{Resources and reproducibility}
\label{sec:native-resources}

\paragraph{Fitting schedules.}
TS trains for 20 epochs, matching the default training length in its source
configuration. In this study, ConvLSTM, PredRNN and SimVP each train for 100 epochs with
batch size four and no early stopping. Further fitting details appear in
the task sections above.
ViBR-WM uses four MCMC chains per regression: each chain runs for 4,000 iterations
in PointMaze and Wall, 48,000 in PhenoCam and 24,000 in SKIPP'D, discarding the first
2,000, 16,000 and 4,000 iterations, respectively. Posterior predictions
combine the retained samples. Neural epochs count passes through training
data; MCMC iterations update posterior samples.

\paragraph{Observed fitting time.}
Table~\ref{tab:native-fitting-time} reports mean elapsed times from the
recorded fitting jobs. The timed jobs include their validation and development scoring
operations where these are part of the fitting job.

\begin{table}[htbp]
\centering\small
\caption{Mean elapsed fitting-stage time in minutes. PhenoCam is
per site, with all four horizons included in each statistical fit.}
\label{tab:native-fitting-time}
\begin{tabular}{@{}lrrrrr@{}}
\toprule
Task & ViBR-WM & TS & ConvLSTM & PredRNN & SimVP\\
\midrule
PointMaze & 80.37 & 8.30 & 9.23 & 68.94 & 23.76\\
Wall & 30.89 & 5.07 & 3.52 & 23.79 & 8.65\\
PhenoCam & 5.36 & 18.84 & 2.97 & 11.18 & 5.05\\
SKIPP'D & 267.44 & 289.82 & 182.65 & 1393.96 & 119.35\\
\bottomrule
\end{tabular}
\end{table}

The SKIPP'D statistical time covers all four horizons and averages the nine
repetitions run after specification selection; the first selected fit's
horizon jobs were performed separately during model selection.
The other entries average ten fits per task, or twenty site-specific fits
for PhenoCam.

%% file: generated/wall_native_final_supplement.tex
\begin{table}[htbp]
\centering
\small
\setlength{\tabcolsep}{5pt}
\begin{tabular}{lrrrr}
\toprule
Method & Frame 15 & Frame 25 & Frame 35 & Frame 45 \\
\midrule
ViBR-WM & 0.1197 & 0.2880 & 0.4045 & 0.4834 \\
Temporal Straightening & 0.2965 & 0.3767 & 0.4751 & 0.5552 \\
ConvLSTM & 0.7135 & 0.8363 & 1.0050 & 1.1157 \\
PredRNN & 0.6338 & 0.9056 & 0.9459 & 1.0108 \\
SimVP & 0.6271 & 0.8024 & 0.9946 & 1.0977 \\
\bottomrule
\end{tabular}
\caption{Wall per-target NRMSE averaged equally over the same 50 trajectories and ten fits. Within each target and trajectory, squared standardized error is averaged over two coordinates before taking its square root. The overall score takes RMS jointly over horizons and coordinates within each trajectory.}
\label{tab:wall-native-horizons}
\end{table}
\begin{table}[htbp]
\centering
\small
\setlength{\tabcolsep}{4pt}
\begin{tabular}{lrrrrr}
\toprule
Repetition & ViBR-WM & Temporal Straightening & ConvLSTM & PredRNN & SimVP \\
\midrule
1 & 0.379430 & 0.477133 & 1.054874 & 0.968377 & 0.902873 \\
2 & 0.369477 & 0.431232 & 0.972504 & 0.899301 & 0.848144 \\
3 & 0.345590 & 0.444623 & 0.948536 & 0.944810 & 0.962504 \\
4 & 0.364094 & 0.405766 & 0.964438 & 0.947825 & 0.961308 \\
5 & 0.357922 & 0.458497 & 1.127643 & 0.905311 & 1.007133 \\
6 & 0.358408 & 0.459781 & 0.909451 & 0.937824 & 1.086151 \\
7 & 0.369363 & 0.448966 & 1.018600 & 0.957266 & 0.941449 \\
8 & 0.360677 & 0.516367 & 0.983907 & 0.977254 & 0.870355 \\
9 & 0.371631 & 0.511543 & 0.911217 & 0.944011 & 0.896056 \\
10 & 0.362972 & 0.436534 & 0.889572 & 0.822778 & 1.061850 \\
\midrule
Sample SD & 0.009275 & 0.034638 & 0.073065 & 0.045080 & 0.079145 \\
\bottomrule
\end{tabular}
\caption{All ten Wall fit aggregates. Each cell averages whole-trajectory RMS scores; sample SD uses denominator nine. ViBR-WM and Temporal Straightening share the trained representation within each repetition; RGB models are initialized and trained independently.}
\label{tab:wall-native-repeats}
\end{table}
\begin{table}[htbp]
\centering
\small
\setlength{\tabcolsep}{5pt}
\begin{tabular}{llrr}
\toprule
Comparator & Target & Difference & 95\% interval \\
\midrule
Temporal Straightening & Overall & -0.0951 & $[-0.1554,-0.0350]$ \\
 & Frame 15 & -0.1767 & $[-0.2183,-0.1309]$ \\
 & Frame 25 & -0.0887 & $[-0.1443,-0.0276]$ \\
 & Frame 35 & -0.0706 & $[-0.1510,0.0084]$ \\
 & Frame 45 & -0.0717 & $[-0.1624,0.0214]$ \\
ConvLSTM & Overall & -0.6141 & $[-0.7168,-0.5119]$ \\
 & Frame 15 & -0.5938 & $[-0.7177,-0.4827]$ \\
 & Frame 25 & -0.5483 & $[-0.6708,-0.4231]$ \\
 & Frame 35 & -0.6005 & $[-0.7566,-0.4396]$ \\
 & Frame 45 & -0.6323 & $[-0.7539,-0.5029]$ \\
PredRNN & Overall & -0.5665 & $[-0.6787,-0.4556]$ \\
 & Frame 15 & -0.5141 & $[-0.6396,-0.3960]$ \\
 & Frame 25 & -0.6176 & $[-0.7439,-0.4888]$ \\
 & Frame 35 & -0.5414 & $[-0.6969,-0.3825]$ \\
 & Frame 45 & -0.5274 & $[-0.6557,-0.3906]$ \\
SimVP & Overall & -0.5898 & $[-0.6968,-0.4810]$ \\
 & Frame 15 & -0.5074 & $[-0.5954,-0.4183]$ \\
 & Frame 25 & -0.5145 & $[-0.6365,-0.3973]$ \\
 & Frame 35 & -0.5901 & $[-0.7401,-0.4333]$ \\
 & Frame 45 & -0.6143 & $[-0.7842,-0.4458]$ \\
\bottomrule
\end{tabular}
\caption{Wall overall and per-target ViBR-WM-minus-comparator error differences and 95\% percentile intervals from 2,000 bootstrap resamples, without adjustment for multiple comparisons. Each resample uses the same selected trajectories across methods and horizons; ViBR-WM and Temporal Straightening resample paired fits that share a trained visual representation, while RGB fits are resampled independently. Conditioning assumptions are given in Section~\ref{sec:representation}.}
\label{tab:wall-native-intervals}
\end{table}
\begin{samepage}
\paragraph{Posterior predictive interval coverage.}
Intervals are computed directly from posterior predictive draws without subsequent recalibration. Empirical coverage is the percentage of observed target coordinates inside their predicted intervals; the nominal level is the probability assigned to an interval by the predictive distribution. At nominal 50\%, coverage is 42.50\% with mean standardized width 0.39678. At nominal 80\%, coverage is 70.50\% with mean standardized width 0.75516. At nominal 95\%, coverage is 89.03\% with mean standardized width 1.15875. For the 95\% intervals, coverage is 90.70\% at frame 15, 88.20\% at frame 25, 88.00\% at frame 35, 89.20\% at frame 45. The summaries were computed from each fit's 8,000 pooled posterior predictive draws and then averaged across fits. Each interval width is divided by the corresponding coordinate's training sample standard deviation. The 400 observed target coordinates comprise 50 trajectories, four target frames and two position coordinates. The 4,000 coverage comparisons reuse the same 400 target coordinates across ten fits. Empirical coverage is below the nominal level for all three interval levels.
\par\end{samepage}
\FloatBarrier

%% file: generated/pointmaze_native_final_supplement.tex
\begin{table}[htbp]
\centering
\small
\setlength{\tabcolsep}{5pt}
\begin{tabular}{lrrrr}
\toprule
Method & Frame 15 & Frame 25 & Frame 35 & Frame 60 \\
\midrule
ViBR-WM & 0.3342 & 0.4936 & 0.6138 & 0.6362 \\
Temporal Straightening & 0.4293 & 0.5921 & 0.6706 & 0.8003 \\
ConvLSTM & 1.4612 & 3.1474 & 2.8551 & 4.1492 \\
PredRNN & 1.3558 & 1.0120 & 1.0448 & 0.9708 \\
SimVP & 1.2258 & 1.1962 & 1.3989 & 1.6190 \\
\bottomrule
\end{tabular}
\caption{PointMaze per-target NRMSE averaged equally over the same 50 trajectories and ten fits. Within each target and trajectory, squared standardized error is averaged over four coordinates before taking its square root. The overall score takes RMS jointly over horizons and coordinates within each trajectory.}
\label{tab:pointmaze-native-horizons}
\end{table}
\begin{table}[htbp]
\centering
\small
\setlength{\tabcolsep}{4pt}
\begin{tabular}{lrrrrr}
\toprule
Repetition & ViBR-WM & Temporal Straightening & ConvLSTM & PredRNN & SimVP \\
\midrule
1 & 0.538330 & 0.621356 & 3.470498 & 1.132026 & 2.131528 \\
2 & 0.548061 & 0.661925 & 4.247897 & 1.140820 & 1.134285 \\
3 & 0.588117 & 0.659380 & 3.140727 & 1.141516 & 1.161737 \\
4 & 0.566431 & 0.692336 & 2.428698 & 1.136431 & 1.781546 \\
5 & 0.571354 & 0.690617 & 2.307833 & 1.141411 & 1.319354 \\
6 & 0.535430 & 0.652987 & 3.720713 & 1.141336 & 1.147300 \\
7 & 0.546357 & 0.655555 & 3.125462 & 1.133990 & 1.403297 \\
8 & 0.561989 & 0.686015 & 1.913387 & 1.138497 & 1.426985 \\
9 & 0.586709 & 0.692770 & 2.946891 & 1.148397 & 1.204068 \\
10 & 0.558993 & 0.670610 & 3.904047 & 1.152801 & 1.645307 \\
\midrule
Sample SD & 0.018501 & 0.022887 & 0.744941 & 0.006242 & 0.327468 \\
\bottomrule
\end{tabular}
\caption{All ten PointMaze fit aggregates. Each cell averages whole-trajectory RMS scores; sample SD uses denominator nine. ViBR-WM and Temporal Straightening share the trained representation within each repetition; RGB models are initialized and trained independently.}
\label{tab:pointmaze-native-repeats}
\end{table}
\begin{table}[htbp]
\centering
\small
\begin{tabular}{lrr}
\toprule
Comparator & ViBR-WM minus comparator & 95\% interval \\
\midrule
Temporal Straightening & -0.1082 & $[-0.1544,-0.0532]$ \\
ConvLSTM & -2.5604 & $[-2.9868,-2.1363]$ \\
PredRNN & -0.5805 & $[-0.6817,-0.4778]$ \\
SimVP & -0.8754 & $[-1.0927,-0.6840]$ \\
\bottomrule
\end{tabular}
\caption{PointMaze overall ViBR-WM-minus-comparator error differences and 95\% percentile intervals from 2,000 bootstrap resamples, without adjustment for multiple comparisons. Each resample uses the same selected trajectories across methods and horizons; ViBR-WM and Temporal Straightening resample paired fits that share a trained visual representation, while RGB fits are resampled independently. Conditioning assumptions are given in Section~\ref{sec:representation}.}
\label{tab:pointmaze-native-intervals}
\end{table}

%% file: generated/phenocam_native_final_supplement.tex
\begin{table}[htbp]
\centering
\small
\setlength{\tabcolsep}{4pt}
\begin{tabular}{lrrrrr}
\toprule
Repetition & ViBR-WM & Temporal Straightening & ConvLSTM & PredRNN & SimVP \\
\midrule
1 & 0.436729 & 1.140654 & 1.006600 & 1.044145 & 0.857985 \\
2 & 0.436722 & 1.354659 & 0.986479 & 1.049004 & 0.793107 \\
3 & 0.436683 & 0.998471 & 0.991100 & 0.976366 & 1.151777 \\
4 & 0.436687 & 3.403910 & 0.952713 & 0.976920 & 0.829342 \\
5 & 0.436793 & 0.781621 & 0.977743 & 1.104369 & 0.757989 \\
6 & 0.436661 & 4.080768 & 0.982536 & 1.081207 & 1.011120 \\
7 & 0.436634 & 1.105045 & 0.937684 & 0.928292 & 0.836056 \\
8 & 0.436706 & 0.950346 & 0.977011 & 1.207690 & 0.888859 \\
9 & 0.436708 & 1.840769 & 0.925207 & 0.884827 & 0.814924 \\
10 & 0.436680 & 1.272850 & 0.981059 & 1.028491 & 0.857481 \\
\midrule
Sample SD & 0.000043 & 1.128262 & 0.025302 & 0.092718 & 0.117044 \\
\bottomrule
\end{tabular}
\caption{All ten PhenoCam fit aggregates. Each entry averages two fixed sites and four horizons; sample SD uses the ten aggregates with denominator nine. Repetition indices pair sites within a method; methods use independent training randomness. ViBR-WM repeats vary MCMC seeds on fixed DINO features, PCA and design; neural fits vary initialization.}
\label{tab:phenocam-native-repeats}
\end{table}
\begin{table}[htbp]
\centering
\small
\begin{tabular}{lrr}
\toprule
Comparator & 13-week primary & 26-week sensitivity \\
\midrule
Temporal Straightening & $[-1.9527,-0.6770]$ & $[-1.9873,-0.7291]$ \\
ConvLSTM & $[-0.6770,-0.3550]$ & $[-0.6854,-0.4865]$ \\
PredRNN & $[-0.7012,-0.4316]$ & $[-0.6451,-0.4588]$ \\
SimVP & $[-0.7025,-0.2146]$ & $[-0.7276,-0.3794]$ \\
\bottomrule
\end{tabular}
\caption{Unadjusted 95\% bootstrap percentile intervals for overall ViBR-WM-minus-comparator NRMSE; negative values favor ViBR-WM. Each block length uses 2,000 draws. Shared calendar weights preserve joint dependence across sites and horizons; training repeats are sampled independently between methods.}
\label{tab:phenocam-native-block-intervals}
\end{table}

%% file: generated/skippd_native_final_main.tex
\begin{table}[htbp]
\centering
\small
\setlength{\tabcolsep}{4pt}
\begin{tabular}{lrr}
\toprule
Method & Overall NRMSE $\pm$ SD & ViBR-WM reduction (\%) \\
\midrule
ViBR-WM & $0.2180 \pm 0.000011$ & --- \\
Temporal Straightening & $0.5448 \pm 0.097869$ & 59.99 \\
ConvLSTM & $0.5072 \pm 0.047198$ & 57.02 \\
PredRNN & $0.4290 \pm 0.019511$ & 49.20 \\
SimVP & $0.4892 \pm 0.024810$ & 55.45 \\
\bottomrule
\end{tabular}
\caption{SKIPP'D 2019 forecasting comparison. Means equally average four horizons and ten fits; SD is the sample standard deviation of ten fit means. Percentages show how much lower ViBR-WM's mean NRMSE is, relative to each comparator's mean NRMSE. Training and validation use data from 2017 and 2018, respectively.}
\label{tab:skippd-native-final-2019}
\end{table}

%% file: generated/skippd_native_final_supplement.tex
\begin{table}[htbp]
\centering
\small
\setlength{\tabcolsep}{4pt}
\begin{tabular}{lrrrr}
\toprule
Method & 5 min & 15 min & 30 min & 60 min \\
\midrule
ViBR-WM & 0.1583 & 0.2046 & 0.2345 & 0.2744 \\
Temporal Straightening & 0.2471 & 0.4144 & 0.6198 & 0.8979 \\
ConvLSTM & 0.4737 & 0.4807 & 0.4985 & 0.5758 \\
PredRNN & 0.3814 & 0.3688 & 0.4405 & 0.5254 \\
SimVP & 0.4435 & 0.4571 & 0.5004 & 0.5557 \\
\midrule
Test origins & 3124 & 3124 & 3124 & 3124 \\
\bottomrule
\end{tabular}
\caption{Mean NRMSE over ten fits on 2019 forecast starting times at each horizon. For each fit and horizon, RMSE across these times is divided by the corresponding 2017 training-target standard deviation. ViBR-WM uses visual, observed-PV and calendar regressors; Temporal Straightening receives RGB images and PV observations through its visual and proprioceptive interfaces, respectively; ConvLSTM, PredRNN and SimVP use RGB only.}
\label{tab:skippd-native-horizons}
\end{table}
\begin{table}[htbp]
\centering
\small
\setlength{\tabcolsep}{4pt}
\begin{tabular}{lrrrrr}
\toprule
Repetition & ViBR-WM & TS & ConvLSTM & PredRNN & SimVP \\
\midrule
1 & 0.217960 & 0.443109 & 0.505898 & 0.426314 & 0.520961 \\
2 & 0.217946 & 0.653381 & 0.527859 & 0.418804 & 0.433620 \\
3 & 0.217951 & 0.435658 & 0.525488 & 0.401077 & 0.501159 \\
4 & 0.217981 & 0.531924 & 0.473520 & 0.454480 & 0.515727 \\
5 & 0.217958 & 0.487822 & 0.449582 & 0.423499 & 0.495842 \\
6 & 0.217953 & 0.597544 & 0.541185 & 0.451756 & 0.494600 \\
7 & 0.217970 & 0.661232 & 0.536513 & 0.424545 & 0.490562 \\
8 & 0.217952 & 0.560585 & 0.415455 & 0.450111 & 0.491347 \\
9 & 0.217951 & 0.411758 & 0.525122 & 0.401035 & 0.468765 \\
10 & 0.217951 & 0.665060 & 0.571067 & 0.438505 & 0.479305 \\
\midrule
Sample SD & 0.000011 & 0.097869 & 0.047198 & 0.019511 & 0.024810 \\
\bottomrule
\end{tabular}
\caption{All ten fit means, each equally averaging four horizon-specific NRMSEs. Four chains form one ViBR-WM fit; repetitions vary MCMC seeds on fixed training features and design. Neural models are initialized and trained independently. Sample SD uses denominator nine.}
\label{tab:skippd-native-repeats}
\end{table}
\begin{table}[!htbp]
\centering
\small
\setlength{\tabcolsep}{4pt}
\begin{tabular}{lrr}
\toprule
Comparator & Difference & 95\% interval \\
\midrule
Temporal Straightening & -0.3268 & $[-0.3881,-0.2688]$ \\
ConvLSTM & -0.2892 & $[-0.3330,-0.2483]$ \\
PredRNN & -0.2111 & $[-0.2384,-0.1871]$ \\
SimVP & -0.2712 & $[-0.3004,-0.2412]$ \\
\bottomrule
\end{tabular}
\caption{Overall ViBR-WM-minus-comparator NRMSE and percentile intervals from 2,000 bootstrap draws. Calendar-week blocks are shared across methods and horizons; training fits are resampled independently within each method. Negative values favor ViBR-WM. Intervals are unadjusted for multiple comparisons.}
\label{tab:skippd-native-contrasts}
\end{table}

%% file: generated/native_point_ensemble_probability.tex
\begin{table}[htbp]
\centering
\small
\begin{tabular}{lrrrr}
\toprule
Dataset & CRPS & Energy & 95\% cov. (\%) & 95\% width \\
\midrule
PhenoCam & 0.237 & --- & 78.87 & 1.096 \\
PointMaze & 0.302 & 0.737 & 93.66 & 2.220 \\
Wall & 0.209 & 0.331 & 89.03 & 1.159 \\
SKIPP'D & 0.111 & --- & 97.43 & 1.223 \\
\bottomrule
\end{tabular}
\caption{ViBR-WM posterior predictive scores. Each fitted model's posterior predictive distribution is scored first, then the scores are averaged over fits and the task's evaluation units. Scores and interval widths are expressed in training-standard-deviation units; coverage (cov.) is the percentage of observed targets within the central 95\% predictive intervals. For scalar Gcc and PV targets, the energy score equals CRPS and is omitted.}
\label{tab:native-statistical-posterior-probability}
\end{table}

\begin{table}[htbp]
\centering
\small
\setlength{\tabcolsep}{4pt}
\begin{tabular}{llrrrr}
\toprule
Task & Method & CRPS & Energy & 95\% cov. (\%) & 95\% width \\
\midrule
PhenoCam & ViBR-WM & 0.293 & --- & 0.28 & 0.001 \\
PhenoCam & Temporal Straightening & 0.546 & --- & 85.45 & 4.811 \\
PhenoCam & ConvLSTM & 0.607 & --- & 44.39 & 0.879 \\
PhenoCam & PredRNN & 0.460 & --- & 79.86 & 1.740 \\
PhenoCam & SimVP & 0.432 & --- & 67.57 & 1.250 \\
\midrule
PointMaze & ViBR-WM & 0.388 & 0.937 & 15.25 & 0.238 \\
PointMaze & Temporal Straightening & 0.346 & 0.839 & 59.88 & 0.944 \\
PointMaze & ConvLSTM & 1.755 & 4.388 & 38.12 & 3.337 \\
PointMaze & PredRNN & 0.927 & 2.116 & 4.88 & 0.175 \\
PointMaze & SimVP & 0.860 & 1.941 & 36.38 & 1.778 \\
\midrule
Wall & ViBR-WM & 0.257 & 0.404 & 19.75 & 0.193 \\
Wall & Temporal Straightening & 0.247 & 0.392 & 63.75 & 0.753 \\
Wall & ConvLSTM & 0.604 & 0.936 & 48.25 & 1.321 \\
Wall & PredRNN & 0.681 & 1.059 & 35.25 & 0.666 \\
Wall & SimVP & 0.572 & 0.913 & 53.50 & 1.195 \\
\midrule
SKIPP'D & ViBR-WM & 0.104 & --- & 1.99 & 0.003 \\
SKIPP'D & Temporal Straightening & 0.193 & --- & 83.85 & 1.185 \\
SKIPP'D & ConvLSTM & 0.299 & --- & 42.53 & 0.587 \\
SKIPP'D & PredRNN & 0.196 & --- & 64.32 & 0.636 \\
SKIPP'D & SimVP & 0.275 & --- & 48.35 & 0.557 \\
\bottomrule
\end{tabular}
\caption{Between-fit dispersion, represented by one equal-weight empirical distribution of ten predicted means per method. CRPS, energy scores and interval widths are expressed in training-standard-deviation units; coverage (cov.) is the percentage of observed targets within the central 95\% empirical intervals. PhenoCam averages equally over horizons and sites, PointMaze and Wall over trajectories and horizons, and SKIPP'D over forecast starting times and horizons. The energy score is omitted for scalar Gcc and PV because it equals CRPS. Central 95\% intervals use linearly interpolated empirical quantiles.}
\label{tab:native-point-ensemble-probability}
\end{table}

%% file: references_arxiv.bib
@misc{lecun2022path,
  title={A Path Towards Autonomous Machine Intelligence},
  author={LeCun, Yann},
  year={2022},
  note={Version 0.9.2},
  url={https://openreview.net/forum?id=BZ5a1r-kVsf}
}

@inproceedings{ha2018worldmodels,
  title={Recurrent World Models Facilitate Policy Evolution},
  author={Ha, David and Schmidhuber, J{\"u}rgen},
  booktitle={Advances in Neural Information Processing Systems},
  volume={31},
  publisher={Curran Associates, Inc.},
  year={2018},
  url={https://papers.nips.cc/paper/2018/hash/2de5d16682c3c35007e4e92982f1a2ba-Abstract.html}
}

@inproceedings{hafner2019planet,
  title={Learning Latent Dynamics for Planning from Pixels},
  author={Hafner, Danijar and Lillicrap, Timothy and Fischer, Ian and
    Villegas, Ruben and Ha, David and Lee, Honglak and Davidson, James},
  booktitle={Proceedings of the 36th International Conference on Machine Learning},
  series={Proceedings of Machine Learning Research},
  volume={97},
  pages={2555--2565},
  publisher={PMLR},
  year={2019},
  url={https://proceedings.mlr.press/v97/hafner19a.html}
}

@inproceedings{hafner2021dreamerv2,
  title={Mastering {Atari} with Discrete World Models},
  author={Hafner, Danijar and Lillicrap, Timothy and Norouzi, Mohammad and Ba, Jimmy},
  booktitle={International Conference on Learning Representations},
  year={2021},
  url={https://openreview.net/forum?id=0oabwyZbOu}
}

@inproceedings{alonso2024diamond,
  title={Diffusion for World Modeling: Visual Details Matter in {Atari}},
  author={Alonso, Eloi and Jelley, Adam and Micheli, Vincent and
    Kanervisto, Anssi and Storkey, Amos and Pearce, Tim and Fleuret, Fran{\c{c}}ois},
  booktitle={Advances in Neural Information Processing Systems},
  volume={37},
  pages={58757--58791},
  year={2024},
  doi={10.52202/079017-1873},
  url={https://proceedings.neurips.cc/paper_files/paper/2024/hash/6bdde0373d53d4a501249547084bed43-Abstract-Conference.html}
}

@inproceedings{dosovitskiy2021vit,
  title={An Image is Worth 16x16 Words: Transformers for Image Recognition at Scale},
  author={Dosovitskiy, Alexey and Beyer, Lucas and Kolesnikov, Alexander and
    Weissenborn, Dirk and Zhai, Xiaohua and Unterthiner, Thomas and
    Dehghani, Mostafa and Minderer, Matthias and Heigold, Georg and
    Gelly, Sylvain and Uszkoreit, Jakob and Houlsby, Neil},
  booktitle={International Conference on Learning Representations},
  year={2021},
  url={https://openreview.net/forum?id=YicbFdNTTy}
}

@inproceedings{caron2021dino,
  title={Emerging Properties in Self-Supervised Vision Transformers},
  author={Caron, Mathilde and Touvron, Hugo and Misra, Ishan and
    J{\'e}gou, Herv{\'e} and Mairal, Julien and Bojanowski, Piotr and Joulin, Armand},
  booktitle={Proceedings of the IEEE/CVF International Conference on Computer Vision},
  pages={9650--9660},
  year={2021},
  url={https://openaccess.thecvf.com/content/ICCV2021/html/Caron_Emerging_Properties_in_Self-Supervised_Vision_Transformers_ICCV_2021_paper.html}
}

@article{scott2014predicting,
  title={Predicting the Present with {Bayesian} Structural Time Series},
  author={Scott, Steven L. and Varian, Hal R.},
  journal={International Journal of Mathematical Modelling and Numerical Optimisation},
  volume={5},
  number={1--2},
  pages={4--23},
  year={2014},
  doi={10.1504/IJMMNO.2014.059942},
  url={https://www.inderscience.com/info/e_inarticle.php?artid=59942}
}

@article{george1993variable,
  title={Variable Selection via {Gibbs} Sampling},
  author={George, Edward I. and McCulloch, Robert E.},
  journal={Journal of the American Statistical Association},
  volume={88},
  number={423},
  pages={881--889},
  year={1993},
  doi={10.1080/01621459.1993.10476353},
  url={https://doi.org/10.1080/01621459.1993.10476353}
}

@article{lusch2018deep,
  title={Deep Learning for Universal Linear Embeddings of Nonlinear Dynamics},
  author={Lusch, Bethany and Kutz, J. Nathan and Brunton, Steven L.},
  journal={Nature Communications},
  volume={9},
  pages={4950},
  year={2018},
  doi={10.1038/s41467-018-07210-0},
  url={https://www.nature.com/articles/s41467-018-07210-0}
}

@article{champion2019coordinates,
  title={Data-Driven Discovery of Coordinates and Governing Equations},
  author={Champion, Kathleen and Lusch, Bethany and Kutz, J. Nathan and Brunton, Steven L.},
  journal={Proceedings of the National Academy of Sciences},
  volume={116},
  number={45},
  pages={22445--22451},
  year={2019},
  doi={10.1073/pnas.1906995116},
  url={https://doi.org/10.1073/pnas.1906995116}
}

@article{gneiting2007proper,
  title={Strictly Proper Scoring Rules, Prediction, and Estimation},
  author={Gneiting, Tilmann and Raftery, Adrian E.},
  journal={Journal of the American Statistical Association},
  volume={102},
  number={477},
  pages={359--378},
  year={2007},
  doi={10.1198/016214506000001437},
  url={https://doi.org/10.1198/016214506000001437}
}

@article{vehtari2021rank,
  title={Rank-Normalization, Folding, and Localization: An Improved
    {$\widehat{R}$} for Assessing Convergence of {MCMC} (with Discussion)},
  author={Vehtari, Aki and Gelman, Andrew and Simpson, Daniel and
    Carpenter, Bob and B{\"u}rkner, Paul-Christian},
  journal={Bayesian Analysis},
  volume={16},
  number={2},
  pages={667--718},
  year={2021},
  doi={10.1214/20-BA1221},
  url={https://doi.org/10.1214/20-BA1221}
}

@article{oquab2024dinov2,
  title={{DINO}v2: Learning Robust Visual Features without Supervision},
  author={Oquab, Maxime and Darcet, Timoth{\'e}e and Moutakanni, Th{\'e}o and Vo, Huy V. and Szafraniec, Marc and Khalidov, Vasil and Fernandez, Pierre and Haziza, Daniel and Massa, Francisco and El-Nouby, Alaaeldin and Assran, Mido and Ballas, Nicolas and Galuba, Wojciech and Howes, Russell and Huang, Po-Yao and Li, Shang-Wen and Misra, Ishan and Rabbat, Michael and Sharma, Vasu and Synnaeve, Gabriel and Xu, Hu and J{\'e}gou, Herv{\'e} and Mairal, Julien and Labatut, Patrick and Joulin, Armand and Bojanowski, Piotr},
  journal={Transactions on Machine Learning Research},
  issn={2835-8856},
  year={2024}
}

@inproceedings{zhou2025dinowm,
  title={{DINO-WM}: World Models on Pre-trained Visual Features Enable Zero-shot Planning},
  author={Zhou, Gaoyue and Pan, Hengkai and LeCun, Yann and Pinto, Lerrel},
  booktitle={Proceedings of the 42nd International Conference on Machine Learning},
  series={Proceedings of Machine Learning Research},
  volume={267},
  pages={79115--79135},
  publisher={PMLR},
  url={https://proceedings.mlr.press/v267/zhou25t.html},
  year={2025}
}

@inproceedings{wang2026straightening,
  title={Temporal Straightening for Latent Planning},
  author={Wang, Ying and Bounou, Oumayma and Zhou, Gaoyue and Balestriero, Randall and Rudner, Tim G. J. and LeCun, Yann and Ren, Mengye},
  booktitle={Proceedings of the 43rd International Conference on Machine Learning},
  series={Proceedings of Machine Learning Research},
  url={https://openreview.net/forum?id=Ik1mKtUYlZ},
  year={2026}
}

@article{proctor2016dmdc,
  title={Dynamic Mode Decomposition with Control},
  author={Proctor, Joshua L. and Brunton, Steven L. and Kutz, J. Nathan},
  journal={SIAM Journal on Applied Dynamical Systems},
  volume={15},
  number={1},
  pages={142--161},
  year={2016},
  doi={10.1137/15M1013857},
  url={https://doi.org/10.1137/15M1013857}
}

@inproceedings{vondrick2016anticipating,
  title={Anticipating Visual Representations from Unlabeled Video},
  author={Vondrick, Carl and Pirsiavash, Hamed and Torralba, Antonio},
  booktitle={Proceedings of the IEEE Conference on Computer Vision and Pattern Recognition},
  pages={98--106},
  year={2016}
}

@article{gao2024bayesianautoencoder,
  title={Bayesian Autoencoders for Data-Driven Discovery of Coordinates, Governing Equations and Fundamental Constants},
  author={Gao, L. Mars and Kutz, J. Nathan},
  journal={Proceedings of the Royal Society A: Mathematical, Physical and Engineering Sciences},
  volume={480},
  number={2286},
  pages={20230506},
  year={2024},
  doi={10.1098/rspa.2023.0506}
}

@article{bardes2024revisiting,
  title={Revisiting Feature Prediction for Learning Visual Representations from Video},
  author={Bardes, Adrien and Garrido, Quentin and Ponce, Jean and Chen, Xinlei and Rabbat, Michael and LeCun, Yann and Assran, Mahmoud and Ballas, Nicolas},
  journal={Transactions on Machine Learning Research},
  year={2024},
  issn={2835-8856},
  url={https://openreview.net/forum?id=QaCCuDfBk2},
  note={Featured Certification}
}

@article{terver2026jepawms,
  title={What Drives Success in Physical Planning with Joint-Embedding Predictive World Models?},
  author={Terver, Basile and Yang, Tsung-Yen and Ponce, Jean and Bardes, Adrien and LeCun, Yann},
  journal={Transactions on Machine Learning Research},
  year={2026},
  url={https://openreview.net/forum?id=cHZn5Gdh8e}
}

@inproceedings{karypidis2025dinoforesight,
  title={{DINO-Foresight}: Looking into the Future with {DINO}},
  author={Karypidis, Efstathios and Kakogeorgiou, Ioannis and Gidaris, Spyridon and Komodakis, Nikos},
  booktitle={Advances in Neural Information Processing Systems},
  volume={38},
  pages={163779--163811},
  publisher={Curran Associates, Inc.},
  year={2025},
  doi={10.52202/085713-5466},
  url={https://proceedings.neurips.cc/paper_files/paper/2025/hash/efaca9631eb6bd5df8f6761c5d9a9fe3-Abstract-Conference.html}
}

@article{brunton2016sindy,
  title={Discovering Governing Equations from Data by Sparse Identification of Nonlinear Dynamical Systems},
  author={Brunton, Steven L. and Proctor, Joshua L. and Kutz, J. Nathan},
  journal={Proceedings of the National Academy of Sciences},
  volume={113},
  number={15},
  pages={3932--3937},
  year={2016},
  doi={10.1073/pnas.1517384113},
  url={https://doi.org/10.1073/pnas.1517384113}
}

@article{brunton2016sindyc,
  title={Sparse Identification of Nonlinear Dynamics with Control ({SINDYc})},
  author={Brunton, Steven L. and Proctor, Joshua L. and Kutz, J. Nathan},
  journal={IFAC-PapersOnLine},
  volume={49},
  number={18},
  pages={710--715},
  year={2016},
  doi={10.1016/j.ifacol.2016.10.249},
  url={https://doi.org/10.1016/j.ifacol.2016.10.249}
}

@article{hirsh2022sparsifying,
  title={Sparsifying Priors for {Bayesian} Uncertainty Quantification in Model Discovery},
  author={Hirsh, Seth M. and Barajas-Solano, David A. and Kutz, J. Nathan},
  journal={Royal Society Open Science},
  volume={9},
  number={2},
  pages={211823},
  year={2022},
  doi={10.1098/rsos.211823},
  url={https://doi.org/10.1098/rsos.211823}
}

@inproceedings{wan2026hauwm,
  author={Wan, Shenghua and Gan, Le and Zhan, De-Chuan},
  booktitle={International Conference on Learning Representations},
  title={Learning to Be Uncertain: Pre-training World Models with Horizon-Calibrated Uncertainty},
  url={https://openreview.net/forum?id=pZuZWRuPyi},
  year={2026}
}

@article{hafner2025dreamerv3,
  title={Mastering Diverse Control Tasks through World Models},
  author={Hafner, Danijar and Pasukonis, Jurgis and Ba, Jimmy and Lillicrap, Timothy},
  journal={Nature},
  volume={640},
  pages={647--653},
  year={2025},
  doi={10.1038/s41586-025-08744-2},
  url={https://doi.org/10.1038/s41586-025-08744-2}
}

@inproceedings{micheli2023iris,
  title={Transformers are Sample-Efficient World Models},
  author={Micheli, Vincent and Alonso, Eloi and Fleuret, Fran{\c{c}}ois},
  booktitle={International Conference on Learning Representations},
  year={2023},
  url={https://openreview.net/forum?id=vhFu1Acb0xb}
}

@inproceedings{shi2015convlstm,
  title={Convolutional {LSTM} Network: A Machine Learning Approach for Precipitation Nowcasting},
  author={Shi, Xingjian and Chen, Zhourong and Wang, Hao and Yeung, Dit-Yan and Wong, Wai-Kin and Woo, Wang-chun},
  booktitle={Advances in Neural Information Processing Systems},
  volume={28},
  pages={802--810},
  publisher={Curran Associates, Inc.},
  year={2015},
  eprint={1506.04214},
  archivePrefix={arXiv},
  url={https://proceedings.neurips.cc/paper_files/paper/2015/hash/07563a3fe3bbe7e3ba84431ad9d055af-Abstract.html}
}

@inproceedings{wang2017predrnn,
  title={{PredRNN}: Recurrent Neural Networks for Predictive Learning Using Spatiotemporal {LSTM}s},
  author={Wang, Yunbo and Long, Mingsheng and Wang, Jianmin and Gao, Zhifeng and Yu, Philip S.},
  booktitle={Advances in Neural Information Processing Systems},
  volume={30},
  pages={879--888},
  publisher={Curran Associates, Inc.},
  year={2017},
  url={https://proceedings.neurips.cc/paper_files/paper/2017/hash/e5f6ad6ce374177eef023bf5d0c018b6-Abstract.html}
}

@inproceedings{gao2022simvp,
  title={{SimVP}: Simpler Yet Better Video Prediction},
  author={Gao, Zhangyang and Tan, Cheng and Wu, Lirong and Li, Stan Z.},
  booktitle={Proceedings of the IEEE/CVF Conference on Computer Vision and Pattern Recognition},
  pages={3170--3180},
  year={2022},
  doi={10.1109/CVPR52688.2022.00317},
  eprint={2206.05099},
  archivePrefix={arXiv},
  url={https://openaccess.thecvf.com/content/CVPR2022/html/Gao_SimVP_Simpler_Yet_Better_Video_Prediction_CVPR_2022_paper.html}
}

@inproceedings{tan2023openstl,
  title={{OpenSTL}: A Comprehensive Benchmark of Spatio-Temporal Predictive Learning},
  author={Tan, Cheng and Li, Siyuan and Gao, Zhangyang and Guan, Wenfei and Wang, Zedong and Liu, Zicheng and Wu, Lirong and Li, Stan Z.},
  booktitle={Advances in Neural Information Processing Systems},
  volume={36},
  pages={69819--69831},
  publisher={Curran Associates, Inc.},
  year={2023},
  doi={10.52202/075280-3060},
  url={https://proceedings.neurips.cc/paper_files/paper/2023/hash/dcbff44d11130e75d09d3930411c23e1-Abstract-Datasets_and_Benchmarks.html}
}

@inproceedings{wang2024predbench,
  title={{PredBench}: Benchmarking Spatio-Temporal Prediction Across Diverse Disciplines},
  author={Wang, ZiDong and Lu, Zeyu and Huang, Di and He, Tong and Liu, Xihui and Ouyang, Wanli and Bai, Lei},
  booktitle={European Conference on Computer Vision},
  series={Lecture Notes in Computer Science},
  volume={15116},
  pages={286--304},
  publisher={Springer},
  year={2024},
  doi={10.1007/978-3-031-73636-0_17},
  url={https://www.ecva.net/papers/eccv_2024/papers_ECCV/html/7567_ECCV_2024_paper.php}
}

@misc{ballou2025phenocamimages,
  title={{PhenoCam Dataset v3.0}: Digital Camera Imagery from the {PhenoCam}
    Network, 2000--2023},
  author={Ballou, K. and Vladich, Z. and Young, A. M. and Milliman, T. and
    Hufkens, K. and Coffey, C. and Begay, K. and Fell, M. and Javadian, M. and
    Post, A. K. and others},
  publisher={ORNL Distributed Active Archive Center},
  year={2025},
  doi={10.3334/ORNLDAAC/2364},
  url={https://doi.org/10.3334/ORNLDAAC/2364}
}

@misc{zimmerman2025phenocamgreenness,
  title={{PhenoCam Dataset v3.0}: Vegetation Phenology from Digital Camera
    Imagery, 2000--2023},
  author={Zimmerman, O. and Young, A. M. and Milliman, T. and Hufkens, K. and
    Ballou, K. and Coffey, C. and Begay, K. and Fell, M. and Javadian, M. and
    Post, A. K. and others},
  publisher={ORNL Distributed Active Archive Center},
  year={2025},
  doi={10.3334/ORNLDAAC/2389},
  url={https://doi.org/10.3334/ORNLDAAC/2389}
}

@article{richardson2018phenocam,
  title={Tracking Vegetation Phenology Across Diverse {North American} Biomes
    Using {PhenoCam} Imagery},
  author={Richardson, Andrew D. and Hufkens, Koen and Milliman, Tom and
    Aubrecht, Donald M. and Chen, Min and Gray, Josh M. and Johnston, Miriam R.
    and Keenan, Trevor F. and Klosterman, Stephen T. and Kosmala, Margaret and
    Melaas, Eli K. and Friedl, Mark A. and Frolking, Steve},
  journal={Scientific Data},
  volume={5},
  pages={180028},
  year={2018},
  doi={10.1038/sdata.2018.28},
  url={https://doi.org/10.1038/sdata.2018.28}
}

@article{nie2023skippd,
  title={{SKIPP'D}: A {SKy Images and Photovoltaic Power Generation Dataset}
    for Short-Term Solar Forecasting},
  author={Nie, Yuhao and Li, Xiatong and Scott, Andea and Sun, Yuchi and
    Venugopal, Vignesh and Brandt, Adam},
  journal={Solar Energy},
  volume={255},
  pages={171--179},
  year={2023},
  doi={10.1016/j.solener.2023.03.043},
  url={https://doi.org/10.1016/j.solener.2023.03.043}
}
